\documentclass[aps,nofootinbib,preprintnumbers,prd,onecolumn,12pt]{article}

\usepackage[T1]{fontenc}                                  
\usepackage{lmodern}                                  
\usepackage{calc}                                 
\usepackage{subfig}
\usepackage{graphicx}                           
\usepackage{booktabs}                 
\usepackage{textcomp}                                 
\usepackage{xspace}                                   
\usepackage{relsize}                                   
\usepackage{amssymb}                               
\usepackage{amsmath}                                  
\usepackage{listings}                                 
\usepackage{microtype}                                  
\usepackage{multirow}                                   
\usepackage{tabularx}                                       
\usepackage{array}                                         
\usepackage{placeins}                                    
\usepackage{cuted}                                              
\usepackage{fixltx2e}
\usepackage{hyperref}

\begin{document}
\begin{titlepage}

\begin{flushright}
{ \bf IFJ-PAN-IV-2019-17 \\
 \today }
\end{flushright}

\vspace{0.1cm}
\begin{center}
{\Huge \bf TAUOLA update for decay channels with $e^+e^-$ pairs in the final state.}
\end{center}
\vspace*{1mm}
\begin{center}
{\bf S.~Antropov$^{a}$, Sw.~Banerjee$^{b}$, Z.~Was$^{a}$, J.~Zaremba$^{a}$}\\
{\em $^a$ Institute of Nuclear Physics Polish Academy of Sciences , PL-31342 Krakow, Poland}\\
{\em $^b$ University of Louisville, Louisville, KY 40292, USA}\\
\end{center}
\vspace{.1 cm}
\begin{center}
{\bf   ABSTRACT  }
\end{center}
With the arrival of high luminosity B-factories like the Belle II experiment,
$\tau$ decay measurements have become more precise than ever,
allowing rarer processes to be explored,
and finer details of $\tau$ decays to be studied.
These are important to understand the spectrum of intermediate particles produced in $\tau$ decays.
Therefore Monte Carlo generators, like the {\tt TAUOLA} program, have to facilitate 
precision analysis as well as confront new models
that constantly emerge with the availability of high statistics experimental data.
New decay channels and models may lead to large variation of 
matrix elements size within the available phase space, 
as in the case when $e^+e^-$ pairs are present in final state.
It requires appropriate presampler of the phase space generator and
a proper documentation to help users introduce their own models.
While releasing a new update, it is important for the {\tt TAUOLA} 
Monte Carlo library
to maintain the general structure of the previous versions to 
preserve backward compatibility.
The aim is to minimize changes from the user perspective.
This paper presents a demonstration
of new models implementation facilitated by the current update.


\vfill %
\vspace{0.1 cm}

\vspace*{1mm}
\bigskip

\end{titlepage}

{\bf PROGRAM SUMMARY}

\begin{small}
\noindent
{\em Program Title: tauola-bbb}                                          \\
{\em Licensing provisions: GPLv2,} 
https://www.gnu.org/licenses/old-licenses/gpl-2.0.en.html                                \\
{\em Programming language: {\tt FORTRAN/C++} }                                   \\
{\em Supplementary material:} TAR BALL with benchmark results, example of the run and README file is localized in directory ......                                \\
{\em Journal reference of previous version: {\em Comput. Phys. Comm.}, 232:220--236, 2018}                  \\
{\em Does the new version supersede the previous version?:} No   \\
{\em Reasons for the new version:} Meeting the needs of present day experiments.\\
{\em \bf  Summary of revisions:}\\
  Phase space presamplers enriched with the possibility to generate narrow peaks
  of $e^+e^-$ pairs, if present in matrix elements of
  4- and 5-body decay channels and  due to electromagnetic or New Physics
  interactions. The examples of physical models are implemented for:
  $\tau^-\to\pi^- e^-e^+\nu_\tau$,
   $\tau^-\to\bar{\nu}_\mu\mu^- e^-e^+\nu_\tau$ and 
   $\tau^-\to\bar{\nu}_e e^- e^-e^+\nu_\tau$ channels, which  are used
   for the collected new algorithm validations.\\
{\em Nature of problem:}
Present day experiments, in particular Belle II, have the capability to measure extremely rare, not yet measured $\tau$ decays,
as well as precise measurements of known decay channels.
They require a Monte Carlo generator capable of generating all desired decays.
They also need the possibility to modify and test new models for such decays.
At the same time the default initialization of the event generator should contain 
decent modeling of known $\tau$ decays.
\\
{\em Solution method:}
The new version supplements previous publications on  {\tt TAUOLA}.
Phase-space presamplers and  new decay channels with additional light lepton
pairs are introduced.
\\
{\em TAR BALL of the future versions of the program may become available  through:} \\
\\

\end{small}

\section{Introduction}

Thanks to its almost 30 years of long history, the {\tt TAUOLA}~\cite{Jadach:1993hs, Chrzaszcz:2016fte}
Monte Carlo (MC) program maintains a large user community while
facilitating present day experiments.
Many solutions could not have been foreseen at the time of creation of the {\tt TAUOLA} program.
As users get accustomed to existing solutions
the original structure of the program needs to be maintained, until an infrastructural update provides a major benefit.
The high statistics data on $\tau$ decays to be collected by the Belle II experiment brings new possibilities and challenges
that nonetheless call for software adaptations.

One of the new aspects of $\tau$ decay analysis is to search for
rare Lepton Flavor Violating (LFV) decays.
For this reason channels with photons and $e^+e^-$ pairs are of interest,
as they constitute the main source of Standard Model (SM) backgrounds to the LFV processes.
This requires the presampler currently used in the phase space generation of {\tt TAUOLA} program
to be extended to accommodate for generation of new channels, in particular with $e^+e^-$ pairs, where sharp peaks are present.

At the same time, old techniques~\cite{Jadach:1993hs}
used to design experimental observable need to remain available. 
The MC generator must continue to be able to work in narrow width approximation 
for intermediate resonances, to ensure compatibility with analytical calculations.

It should be noted that in order to maintain the overall structure of the project,
we have refrained from making drastic modifications e.g.
ones that would change structure of specific arguments used by different functions and routines.
This may seem to be sub-optimal from Information Technology point of view 
(especially if one ignores requirements of other programs),
but can be justified from the point of view of physics complexity.

In this paper, we have documented present software structure of the {\tt TAUOLA}
package for simulation of the $\tau$-lepton decay deployed with the new infrastructural 
development of efficient presamplers for phase-space generation. We have
re-organized storage of the data to make basic elements (routines) encapsulated,
and ready for example to be used as instants once the code is re-written
piece-by-piece to C. Also for that purpose, the common blocks were
separated into two groups. The ones which contain constants like $\pi=3.1415...$,
the fine structure constant $\alpha_{QED}$, the meson decay constants $f_\pi$, etc.
and the ones which collect information of the simulated sample.
The latter one are addressed in restricted form
and are ready to be replaced by arrangements suitable for parallelization of the code.
Since the re-write of the whole code to C must be completed first,
the actual solution is not complete.

Specifics of $\tau$-lepton decays are phenomenological aspect of parameterization
of its decay channels matrix elements or of its hadronic currents.
For the same decay channel, several variants of decay matrix elements or
hadronic currents may be used in the paper,
which are expected to be replaced in future by the updated variants
when the data from the Belle~II experiment can be used to decide
on the preferred parameterization of matrix elements.

We use decay channel parameterizations with different variants,
some of which have simplifications particularly useful for tests.
Obviously, numerical results will differ with that variants, as can be seen in the paper.

We have addressed the questions of numerical stability for the calculation.
Related problems and solutions appear in evolution of practically every large Monte
Carlo program. Unfortunately this is rarely documented, or even mentioned. 
In Ref.~\cite{Przedzinski:2022byu} we have addressed such testing and development aspects,
using our projects as examples. This seems trivial, but often lead to difficulties.
That is why, we have decided to highlight this acpect.
This can be particularly demanding if during the evolution of the project,
the precision improves significantly.
Transition from the single to double precision serves as an example.
Note that this was more challenging for {\tt PHOTOS} Monte Carlo~\cite{Davidson:2010ew}
(available from \url{http://photospp.web.cern.ch/photospp/}),
where double precision was not enough and
dedicated algorithms preventing accumulations of rounding errors were necessary.

Our paper is organized as follows. 
In section~\ref{phstruct} we explain how the phase space in {\tt TAUOLA} 
is structured and parameterized. Section~\ref{mefeat}
introduces models which are useful for the technical tests of features expected in
bremsstrahlung-like emission of $e^+e^-$ pairs in the final state.
Section~\ref{enhance} describes modifications of 
the phase space presampler required for this type of models with $\tau$ decaying into 5 particles.
Section~\ref{presopt} 
compares the performance of the new modifications of the 
phase space parameterization 
with the test matrix elements described in section~\ref{mefeat}. 
Different sets of initialization parameters for the presampler are discussed.
These tests are important in the context of user modifications
as they demonstrate how to optimize generation of the 
phase space density variation resulting from new models.
It documents the ``red flags'' that we have experienced, and describes the steps to be followed
to resolve similar issues that can appear during phase space optimization by other users.
Section~\ref{doublep} describes technical tests of the phase space presamplers.
In section~\ref{exampl}, we present examples of physical models implemented for:
\begin{itemize}
\item $\tau^-\to\pi^- e^-e^+\nu_\tau$,
\item $\tau^-\to\bar{\nu}_\mu\mu^- e^-e^+\nu_\tau$, \hskip 10mm
$\tau^-\to\bar{\nu}_e e^- e^-e^+\nu_\tau$.
\end{itemize}
Selected plots and branching ratios (BR) calculated with the {\tt TAUOLA} Monte Carlo are also presented.
Section~\ref{sum} summarizes the paper.

Some technical aspects of the paper are delegated to Appendices. 
Extended documentation of the presamplers may be required by the users.
It is provided in the Appendix~\ref{sec:presamplers}.
This is necessary since a lot of work on new models is 
expected from the growing user community.
Introduction of decay channels with $e^+e^-$ pairs in the final state,
requires not only extension of the presamplers 
but also new tests of technical and scientific nature. 
Technicalities are explained in Appendix~\ref{sergiej}, 
but tests results are spread over
Sections~\ref{mefeat},~\ref{doublep} and~\ref{exampl}.

\section{Phase space generator structure}\label{phstruct}

In the {\tt TAUOLA} program,
phase space generation and matrix element calculations are performed in separate units.
Event construction starts from phase space. 
Through change of 
variables it can be optimized to generate more events in the regions
where peaks of the matrix elements are expected.
Finally, from the phase space Jacobian ($J$) 
and the matrix element (${\cal M}$) a decay is simulated.
The decay width is obtained using the canonical calculation:
\begin{equation}
 d\Gamma=W*\mathlarger{\prod_i^N dx_i},\label{dgam}
\end{equation}
where $\{x_i\}$ denote set of random numbers and weight:

\begin{equation}
 W=\frac{|{\cal M}|^2*J}{2 M_\tau}\label{weight},
\end{equation}
where $M_\tau$ is the $\tau$ lepton mass.

If weight is bigger than the random number multiplied by 
the maximum weight,
event is accepted, otherwise it is rejected. 
We assume the reader is familiar with notation described in 
Ref.~\cite{Jadach:1993hs}.

Exact functional form of phase space Jacobians depends on the number 
of final state particles and parameterization chosen. 
Phase spaces are usually parameterized using Lorentz invariant quantities and angles of outgoing particles. In {\tt TAUOLA}, the square of the invariant masses
of systems of particles with descending number of decay products are used as the Lorentz invariant quantities\footnote{
For full functional forms we delegate reader to~\cite{Jadach:1993hs}.
Here we would like to concentrate only on parts modifiable by the 
{\tt TAUOLA} user through available tools for user re-definitions.
}.
For some matrix elements calculation, special features 
are expected in invariant masses like Breit-Wigner peaks.
If no significant features are expected in the invariant mass squared $s$, 
the relevant coordinate of phase space is translated 
into a random number in [0,1] range using:
\begin{equation}
s=s_{min}+(s_{max}-s_{min})\cdot x.
\label{flatps}
\end{equation}
Let us call Eq.~\eqref{flatps} a flat type coordinate presampler. 
If the matrix element contains a resonance in the variable $s$, 
we introduce a change of variable:
\begin{align}
 \alpha_{min} & = \arctan{\frac{s_{min}-M_R^2}{\Gamma_R M_R} }, \nonumber\\
 \alpha_{max} & = \arctan{\frac{s_{max}-M_R^2}{\Gamma_R M_R} }, \nonumber\\
  \alpha & = \alpha_{min}+(\alpha_{max}-\alpha_{min})\cdot x, \nonumber \\
  s    & = M_R^2+\Gamma_R M_R \tan\alpha, \label{resps}
\end{align}
where $M_R$ and $\Gamma_R$ are parameters describing the 
resonance present in the matrix element.
Let us call Eq.~\eqref{resps} a resonant type phase space presampler\footnote{
We have checked that it is also convenient to compensate for matrix element
peak of e.g. $\gamma^* \to e^+e^-$ propagator, 
then $M_R=2m_e$ and $\Gamma_R=2m_e$ is used.

Change of variables, in principle,
does not affect simulation results. It affects rejection efficiency though.
If the choice is very inappropriate, the execution time of the algorithm
may increase by unacceptable orders of magnitude.
}.
If the resonance of the matrix element contributes with probability $P$ 
to the total decay rate, then a numerically similar $P$ can be used 
to mix phase space 
parameterizations as explained in Ref.~\cite{Jadach:1993hs}.
At the same time, values for $M_R$ and $\Gamma_R$
should be similar to mass and width of the resonance in matrix element.
Therefore, phase space generation has parallel channels that correspond to
the features of matrix element but use of their properties of should\footnote{
Use of resonant type phase space presampler
compensates for peaks in matrix element.} 
not affect resulting distributions. Monte Carlo calculated
partial width and its uncertainty are accurately calculated 
as long as significant amount of over-weighted events are not generated.
If the correspondence is poor, and somewhat different values of $P$, $M_R$ and $\Gamma_R$ are present in the
matrix element calculation than in phase space parameterization,
event generation will suffer from low efficiency\footnote{
Number of accepted events divided by number of generated events.}.
Large number of over-weighted events relative to the sample size and big uncertainty on partial width
are signs that features of matrix element
and phase space generation do not adequately correspond to each other.

Whenever a change of variable is performed, 
the Jacobian of a such change enters into the phase space formula.
In our case we have multiple channels in presampler, which with necessitates multiple change of variables.
This can be taken into account by introducing harmonically averaged Jacobian $(J_{total})$ from $n$ different
channels in the presampler:
\begin{equation}
  \frac{1}{J_{total}}=\mathlarger{\mathlarger{\sum_{i=1}^n}} \frac{P_i}{J_i},
\label{jacobian}
\end{equation}
where $P_i$ is the probability of the $i^{th}$ presampler channel, and
$J_i$ is the Jacobian corresponding to change of variable performed for the $i^{th}$ channel.

We delegate Appendix~\ref{sec:presamplers} to detailed information on
the presampler channels and the parameterization types used.
We would like to stress that presampler parameters need to be 
tailored for each decay channel separately.

The appearance of over-weighted events should be interpreted as warning.
Their presence differently affects the distributions and the calculation of decay rate,
where the actual error printed in the code remain valid,
as long as it remains smaller than what would be expected 
from $N$ accepted events:  $(~1/\sqrt{N})$. For distributions,
especially if one is interested in corners of the phase space, there are no universal rules.
There is nothing new in this respect for the present version of the program.

If  decay channels parameterizations are changed, over-weighted events appear
and results of the simulation may be affected in particular corners of phase-space.
One can fix the problem either with changes of presampler parameters
or with the help of maximum weight which is found from
the maximum of the event weights simulated at the initialization.

\section{Features expected in matrix element}\label{mefeat}

Before introducing any changes to phase space generator we would
like to discus what to expect from channels with bremsstrahlung of pair emission.
For this purpose, we have chosen models
for $\tau^-\to\pi^- e^-e^+\nu_\tau$ as in Ref.~\cite{Guevara:2013wwa} 
and for $\tau^-\to\bar{\nu}_\mu\mu^- e^-e^+\nu_\tau$
we considered the approach of factorizing the matrix element for pair emission.
Both predict dominant contribution from near real photon, manifesting as
extremely large peak in $e^-e^+$ mass spectra.
This can be simplified\footnote{
We opt for simplified model only for testing purposes, 
as it is then clear what we are testing. Otherwise,
introducing physical models right away
might hinder tests in an unpredictable manner.
} to the matrix element:

\begin{equation}
{\cal M}= \frac{1}{k^2},
\label{me1}
\end{equation}
where $k=P_{e^-}+P_{e^+}$.
Acollinearity\footnote{
Confirming that this acollinearity is handled well means that acollinearity of
photon in channels with $\gamma$ in the final state also can be modeled adequately.
} of emitted pair with the third
charged particle in the $\tau$ decay is the next feature to be addressed.
We model this feature with a minor extension to the previous equation:
\begin{equation}
{\cal M}= \frac{1}{k^2}\frac{2P_1}{(2P_1\cdot k+k^2)},
\label{me2}
\end{equation}
where $P_1$ is four momentum of the
third charged particle (either $\pi^-$ or $\mu^-$).
By comparing the above equation to e.g. Eq.~2
of Ref.~\cite{Guevara:2013wwa}, 
one can see that this is a simplified form of
bremsstrahlung only from the final state.
Eq.~\eqref{me1} and Eq.~\eqref{me2}, while being far away from fully describing the complete underlying processes,
are convenient for testing, and will be used in following sections to illustrate the relevant subtle points. 
They can be easily integrated analytically and numerically.
In Appendix~\ref{sergiej} we use similarly simplified models 
(e.g. ${\cal M}= \frac{const.}{k}$) for testing the numerical stability of factorization of the bremsstrahlung part of full matrix element. 
It further validates use of matrix elements presented in this section.

\section{Enhancing phase space presampler for 5-particle decays}\label{enhance}

From the considerations of the previous section, we can conclude that
it is important for the presampler to accommodate
for the peaks in the mass of two particles\footnote{
Ordering of particles is important here.
Last particle is always assumed to be $\nu_\tau$. We choose here to put
$e^-$ and $e^+$ as the $3^{rd}$ and $4^{th}$ daughter.}.
The phase space presampler for 5-particle decays used to have only
two channels, where resonant type parameterization as described by Eq. \eqref{resps}
was used only for the invariant mass of three particles:
($2^{nd}$, $3^{rd}$ and $4^{th}$) or 
($1^{st}$, $3^{rd}$ and $4^{th}$) .
This is insufficient for a decay such as 
$\tau^-\to\bar{\nu}_\mu\mu^- e^-e^+\nu_\tau$, 
which clearly requires resonant type presampler in 
the invariant mass of two particles.

To introduce new phase space channel we internally
defined\footnote{We want to minimize changes to the
phase space generator as well as user interface.
Introducing new arguments for the routine is 
therefore at present undesirable.}
enhancement for both mass and width of resonance, set to\footnote{
It is good enough for the low mass enhancement and
in future can be immediately used for narrow
resonances in other channels.} 0.001 GeV ($\sim 2m_e$).
New channel is activated with probability of 0.8 if
the $3^{rd}$ and the $4^{th}$ particle in the decay are $e^-$ and $e^+$, respectively.
Such choice greatly improved efficiency of generation of demo model as detailed in section~\ref{cmuee}.
Appendix~\ref{p5} documents the new structure of this presampler.

\section{Presampler parameters optimization}\label{presopt}

While adding new decay channels through the interface for user re-definitions
of {\tt TAUOLA}, it is important to keep in mind that presampler parameters
can account for the newly introduced matrix elements.
For our demo model of Eq.~\eqref{me1} that means optimizing the parameters for peak in the low mass region of invariant mass of $e^+e^-$ pair.
In this section we will show how the choice of parameters may influence event generation.
We should note that such choice doesn't need to be perfect
as it affects only the efficiency of event generation.
Usually it is not worth putting in more effort into searching of optimal parameter values
if the efficiency exceeds 10\% and not too many\footnote{
Relative to the sample size and specific matrix elements.
Usually if less than 1\% of the events are over-weighted, it is acceptable.
If it is more, then we suggest double checking the matrix element and
the chosen presampler parameters. 
One should also check if there are any deformations in generated distributions.
} over-weighted events are generated. 
That being said, for complicated channels with high 
multiplicity in final state particles, high efficiencies might be unattainable.

\subsection{Case of $\tau^-\to\bar{\nu}_\mu\mu^- e^-e^+\nu_\tau$}
\label{cmuee}

Prior to introduction of the enhancement described in Section~\ref{enhance},
the structure of phase space presampler for decays into 5 particles
was insufficient and suffered from up to 48\% of over-weighted events.
This happened due to lack of phase space for the
resonant type parameterization in the
invariant mass constructed from four-momenta of two particles.
Historically this presampler was used for channels with 4 hadrons only in the final state.
In such cases, formation of narrow resonance in the mass of two particles was not expected\footnote{
Mass of four hadron system is rather sizable in comparison to the $\tau$ lepton mass.
Thus, there is little room  for the system to be created from the decay of a resonance.
That is why resonance presamplers were prepared only for two scalar or three scalar system from the decays of $\rho$ and $a_1$, respectively.
The first one, could be used for enhancement now.
}.
No known hadronic current featured such resonances at the time.
The need has arisen only now with advent of high precision data from experiments like Belle II, allowing for investigation of
rare decays.

\begin{figure}
\includegraphics[width=0.5\textwidth]{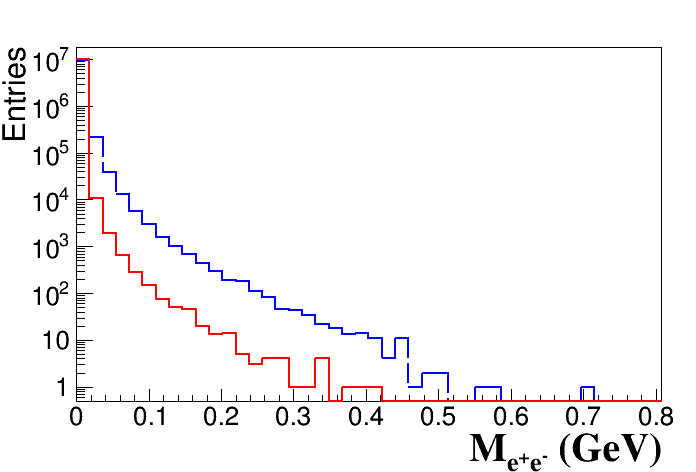}
\includegraphics[width=0.5\textwidth]{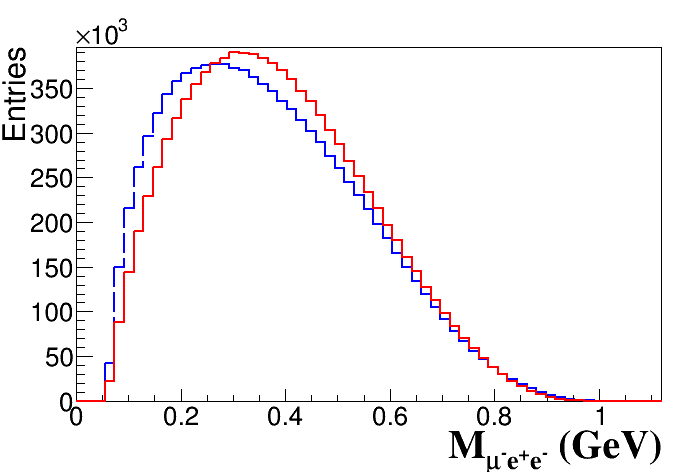}
\includegraphics[width=0.5\textwidth]{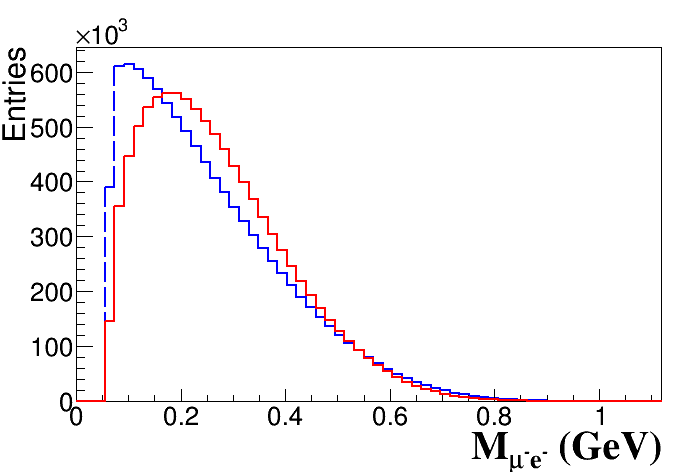}
\includegraphics[width=0.5\textwidth]{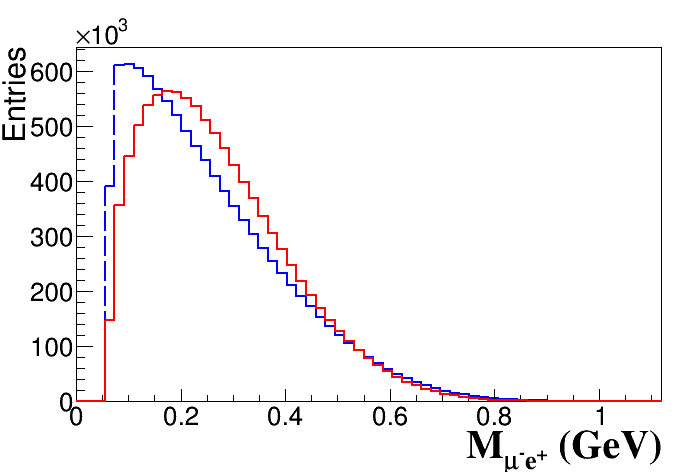}
\caption{
Example plots from {\tt MC-TESTER}~\cite{Golonka:2002rz, Davidson:2008ma}: comparison for
$\tau^-\to\bar{\nu}_\mu\mu^- e^-e^+\nu_\tau$ decay modeled with Eq.~\eqref{me1}
with different phase space parameters, 
as listed in Table~\ref{pre4}.
Invariant masses of $e^+e^-$, $\mu^-e^+e^-$,
$\mu^-e^-$ and $\mu^-e^+$ systems are shown on 
top-right, top-left, bottom-left and bottom-right, respectively.
These plots show an extreme case where the presampler parameters are completely
disconnected from matrix element used, and therefore shapes of the plots are affected.
Dashed blue line represents plots with probability = 0.0,
which is same as using unmodified phase space generator,
while solid red line is obtained while using probability = 0.8.
Large differences in the distributions
come from high amount (48\%) of over-weighted events in sample generated with probability = 0.0.
Samples of 10M events were used for this comparison.}
\label{muee}
\end{figure}

In this subsection we present results from optimization of parameters for the channel newly added to the presampler.
Table~\ref{pre4} collects information on how the probability
of resonance-like phase space enhancement branch
in mass of $e^+e^-$ affects the efficiency of event generation.
At the same time it barely affects the calculated partial width
(see Table~\ref{pre4}), which is how it should be.
Tests were performed on a sample of 10M events.
Figure~\ref{muee} compliments these results with
example plots obtained by comparing the optimized presampler
(probability = 0.8) to the old one (probability = 0.0).
Plots were obtained with the help of {\tt MC-TESTER}~\cite{Golonka:2002rz, Davidson:2008ma}.
The differences are large, exposing the impact from  over-weighted faulty events obtained while using the old presampler.

\begin{table}[h]
\begin{center}
\resizebox{\textwidth}{!}{
\begin{tabular}{|l|l|l|l|l|}
\hline
\cline{1-5}\multicolumn{1}{|c|}{Presampler param.}&\multicolumn{4}{|c|}{ Presampler performance }\\
 \hline
Probability & Efficiency & over-weighted & width [GeV] & relative error \\
 \hline
 0.0 & 0.0005 & 48\% & $0.13784\cdot10^{-2}$ & 0.00204682 \\
 \hline
 0.2 & 0.036 & 0 & $0.13747\cdot10^{-2}$ & 0.00014140 \\
 \hline
 0.4 & 0.066 & 0 & $0.13742\cdot10^{-2}$ & 0.00012545 \\
 \hline
 0.6 & 0.099 & 0 & $0.13744\cdot10^{-2}$ & 0.00011399 \\
 \hline
 0.8 & 0.132 & 0 & $0.13741\cdot10^{-2}$ & 0.00010129 \\
 \hline
\end{tabular}}
\end{center}
\caption{Probability of internally defined phase space channel
and resulting efficiency of generation
for demonstrative $\tau^-\to\bar{\nu}_\mu\mu^- e^-e^+\nu_\tau$ decay
as defined by Eq.~\eqref{me1}.
Tests were done with sample size of 10M events.}
\label{pre4}
\end{table}

\subsection{Case of $\tau^-\to\pi^- e^-e^+\nu_\tau$}\label{cpiee}

In this subsection we want to present procedure for optimizing 
parameters of presampler for decays into 4 particles.
From the matrix element for our demo model (Eq.~\eqref{me1}) and presampler structure
(Appendix~\ref{p4}), we may conclude
that we need modify presampler parameters 
to facilitate the $e^-e^+$ peak with relatively high probability.
Other parameters are of secondary importance and should correspond to roughly flat phase space.
Table~\ref{pre3} collects selected values of 
presampler parameters and their resulting efficiency.
Parameters $P_A$, MA, GA were left at default values\footnote{
That is $P_A=0.0$, therefore MA and GA are irrelevant},
and therefore are not listed in the table.
Without any optimization of presampler the
$\tau^-\to\pi^- e^-e^+\nu_\tau$ channel had
an efficiency of generation at the level of 0.00024,
with over 45\% over-weighted events.
Such results are unacceptable, and values of partial width and its error collected
in Table~\ref{pre3} have little meaning on the shapes of the distributions when so many over-weighted events are present.
Fortunately, in this case, all relevant presampler channels
are available and it is enough to modify parameters 
through interface of {\tt TAUOLA-bbb} via user re-definitions.

\begin{figure}
\includegraphics[width=0.5\textwidth]{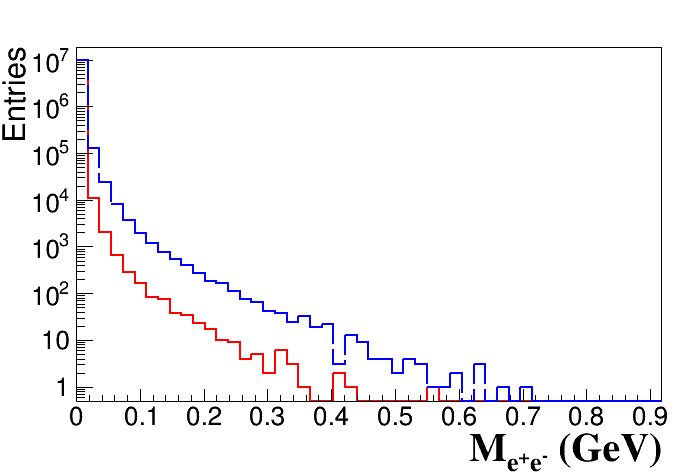}
\includegraphics[width=0.5\textwidth]{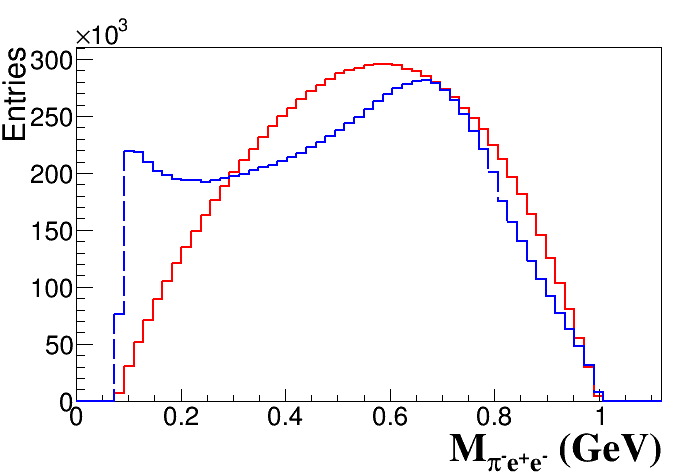}
\includegraphics[width=0.5\textwidth]{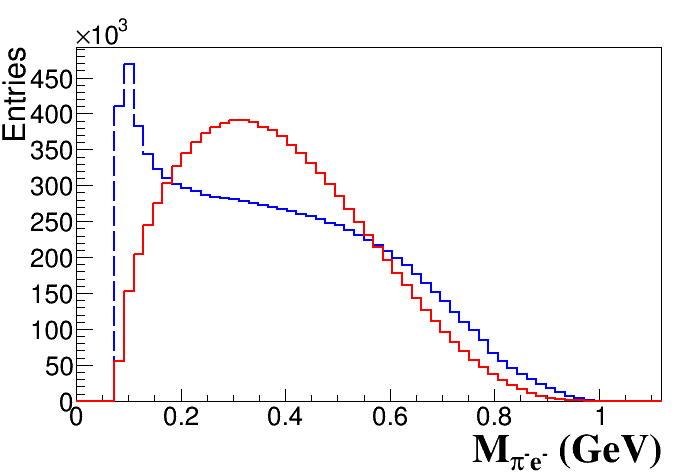}
\includegraphics[width=0.5\textwidth]{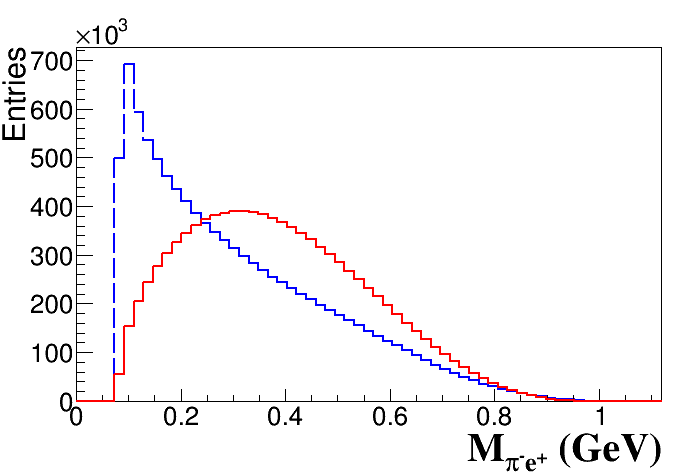}
\caption{Example plots from {\tt MC-TESTER}~\cite{Golonka:2002rz, Davidson:2008ma}: 
comparison for
$\tau^-\to\pi^- e^-e^+ \nu_{\tau}$ decay modeled with Eq.~\eqref{me1}.
Invariant masses of $e^+e^-$, $\pi^-e^+e^-$,
$\pi^-e^-$ and $\pi^-e^+$ systems are shown on 
top-right, top-left, bottom-left and bottom-right, respectively.
Dashed blue line represents plots before any changes to phase space parameters and 
solid red lines show best  set of parameters,  
as listed in Table~\ref{pre3}. 
Large differences in the distributions
come from high number of over-weighted events (45\%) in sample generated with
unmodified phase space parameters.
Samples of 10M events were used for this comparison.
}
\label{piee}
\end{figure}

\begin{figure}
\includegraphics[width=0.5\textwidth]{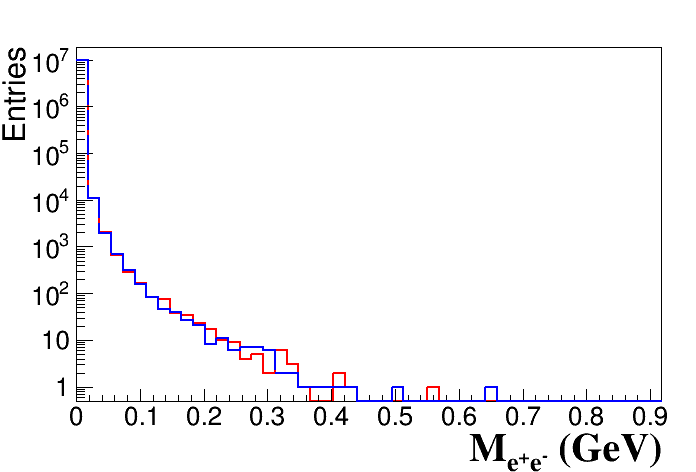}
\includegraphics[width=0.5\textwidth]{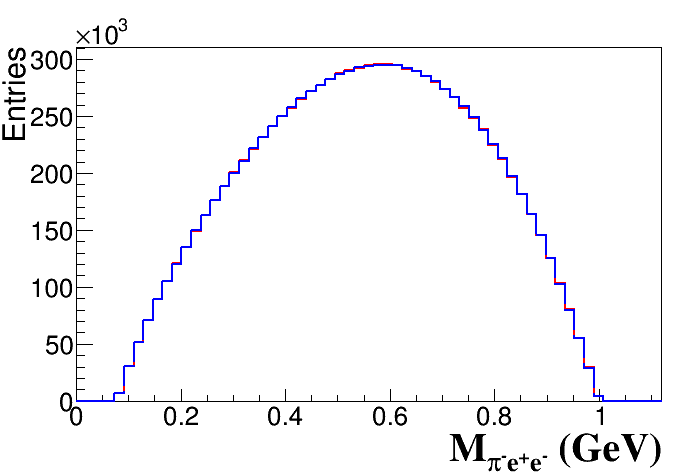}
\includegraphics[width=0.5\textwidth]{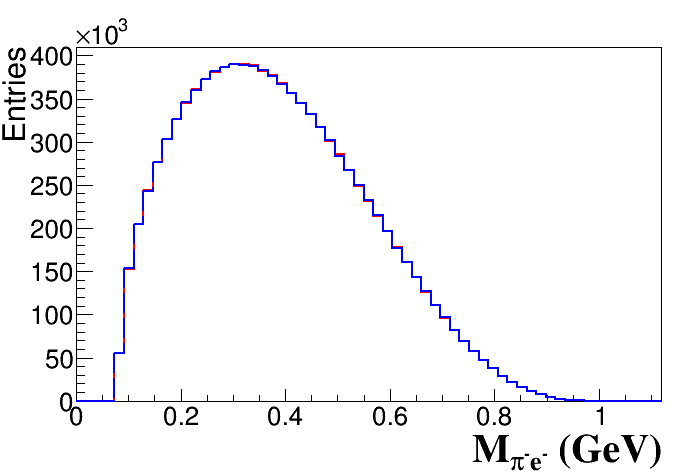}
\includegraphics[width=0.5\textwidth]{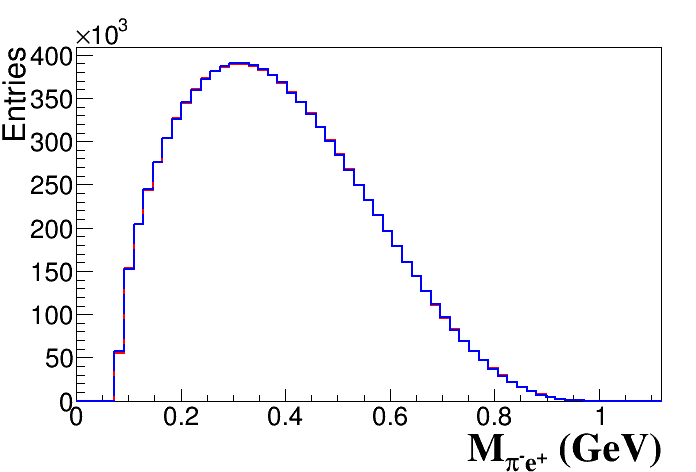}
\caption{Example plots from {\tt MC-TESTER}~\cite{Golonka:2002rz, Davidson:2008ma}: 
comparison for
$\tau^-\to\pi^- e^-e^+ \nu_{\tau}$ decay modeled with Eq.~\eqref{me1}
with different phase space parameters chosen
(see Table~\ref{pre3} for all tested sets of parameters).
Invariant masses of $e^+e^-$, $\pi^-e^+e^-$,
$\pi^-e^-$ and $\pi^-e^+$ systems are shown on 
top-right, top-left, bottom-left and bottom-right, respectively.
Dashed blue line represents plots with sub-optimal phase space parameters choice 
($P_B=0.5$, MX=1.251, GX=0.599, MB=0.001, GB=0.001), while
the solid red line is obtained  with best found set of parameters
($P_B=0.8$, MX=0.9, GX=0.8, MB=0.001, GB=0.001).
Lack of significant differences
for these and all other plots is a proof that presampler works correctly,
that is parameters (as long as they are within reasonable range)
affect only efficiency of generation and do not affect shapes of the distributions.
Samples of 10M events were used for this comparison.}
\label{pieepre}
\end{figure}

Table~\ref{pre3} collects information on how the probability
of branch with resonance-like enhancement of phase phase
in the mass of $e^+e^-$ affects efficiency of generation.
Figure~\ref{piee} compares with the help of
{\tt MC-TESTER}~\cite{Golonka:2002rz, Davidson:2008ma} selected mass distributions\footnote{
{\tt MC-TESTER} compares all possible invariant mass distributions.
We do not think presenting all of them here in beneficent
to the reader, therefore we select only few plots.} obtained with old (first line in Table~\ref{pre3})
and optimized (last line in Table~\ref{pre3})
parameters of the presampler.
Figure~\ref{pieepre} presents similar comparison between
samples obtained with presampler parameters set to 
the second last and the last set of values listed in Table~\ref{pre3}.

\begin{table}[h]
\resizebox{\textwidth}{!}{%
\begin{tabular}{|l|l|l|l|l|l|l|l|l|}
\hline
\cline{1-5}\multicolumn{5}{|c|}{Presampler param.}&\multicolumn{4}{|c|}{Presampler performance}\\
\hline
$P_B$ & MX & GX & MB & GB & eff. & overw.
& width[GeV] & rel.error \\
\hline
 0.0 & 1.251 & 0.599 & 0.7759 & 0.1479 &
 0.00024 & 45\% & 0.616/NaN & 0.2423/NaN \\
\hline
0.5 & 1.251 & 0.599 & 0.001 & 0.001 &
0.0458 & 0\% (14) & 0.424 & 0.0006\\
\hline
0.5 & 1.251 & 0.7 & 0.001 & 0.001 &
0.0519 & 0\% (8)& 0.427/NaN & 0.0035/NaN\\
\hline
0.5 & 1.0 & 0.7 & 0.001 & 0.001 &
0.0912 & 0\% (10) & 0.425 & 0.0001\\
\hline
0.5 & 0.9 & 0.8 & 0.001 & 0.001 &
0.1258 & 0\% (11) & 0.426 & 0.0010\\
\hline
0.8 & 0.9 & 0.8 & 0.001 & 0.001 &
0.1943 & 0\% (1) & 0.425 & 0.0001\\
\hline
\end{tabular}}
\caption{Presampler parameters and resulting efficiency of generation,
over-weighted events, width and relative error of the width
for demo model of $\tau^-\to\pi^- e^-e^+\nu_\tau$ decay.
First line corresponds to initialization as of previous {\tt TAUOLA} versions.
Set of parameters with the best efficiency is considered the optimal one.
Sample size of 10M events was used. 
Unphysical models can sometimes result in NaN value from matrix element calculation.
Even one such unphysical event in generation renders width and its error to be NaN.
In case when NaN are obtained in samples of 10M events, values obtained with a smaller set of 10k sample, which avoids generation of the unphysical event, are listed.
As these are only demonstrative models and such events happen due to sub-optimal parameterization, it is not worth resolving the issue.}
\label{pre3}
\end{table}

The main aspect of optimizing presampler is reducing number of
over-weighted events, as these hinder calculation of partial
width along with its error, and in extreme cases alter shapes of distributions. 
With sample of 10k events we obtain
partial width of 0.616 GeV and relative error of 0.242
for the old presampler, while after optimization of the parameters we get width of 0.425 GeV and error of 0.003 with same sample size.
Note that investigated demonstrative model is unphysical
and so is obtained partial width.
The take-away message from this exercise is the impact of presampler parameters on error estimation.
With the results collected so far, we can conclude that narrow width resonance structure which we used in phase space enhancement is sufficient to handle the logarithmic structure
of $e^-e^+$ pseudo-singularity.
This is not unexpected because the propagator of resonance: $1/(s+M^2)$
in the limit of $s\gg M^2$ resembles the propagator $1/s$ in our demo model.

\subsection{Investigating acollinearity of emitted pair}\label{coll}

\begin{figure}
\includegraphics[width=0.5\textwidth]{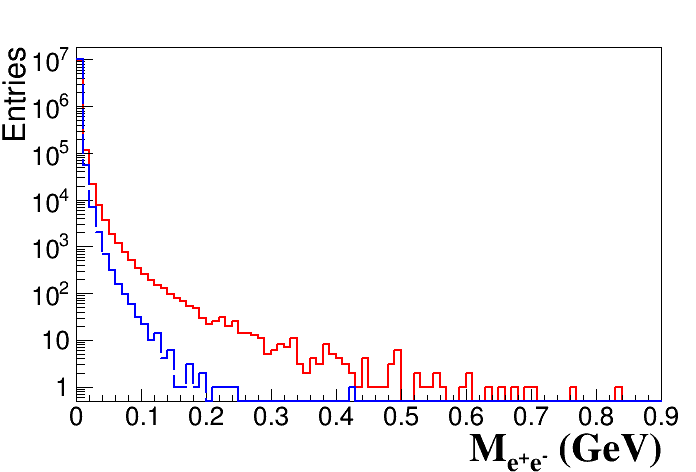}
\includegraphics[width=0.5\textwidth]{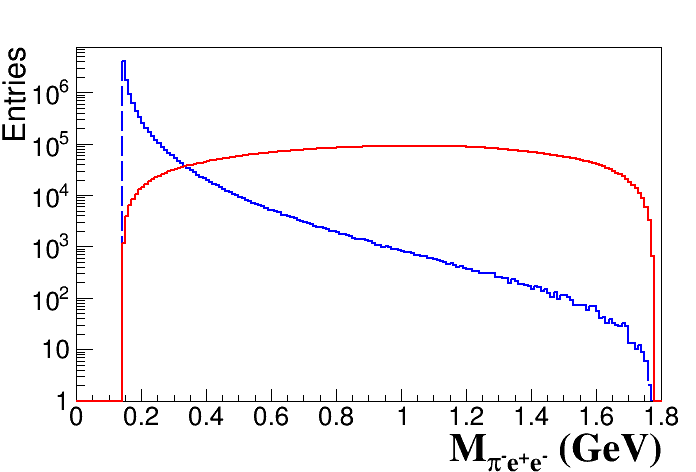}
\includegraphics[width=0.5\textwidth]{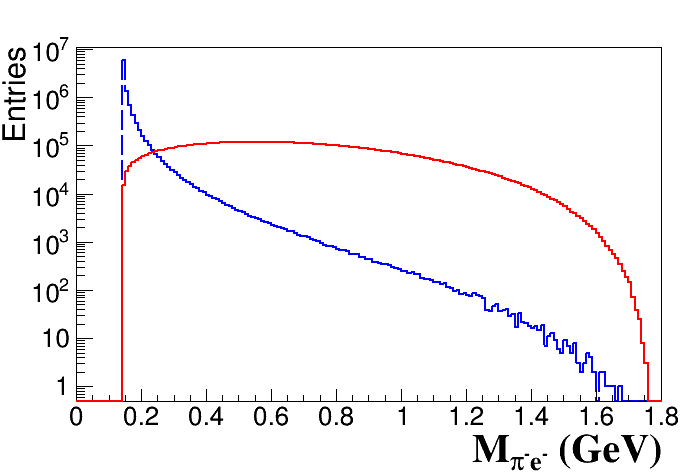}
\includegraphics[width=0.5\textwidth]{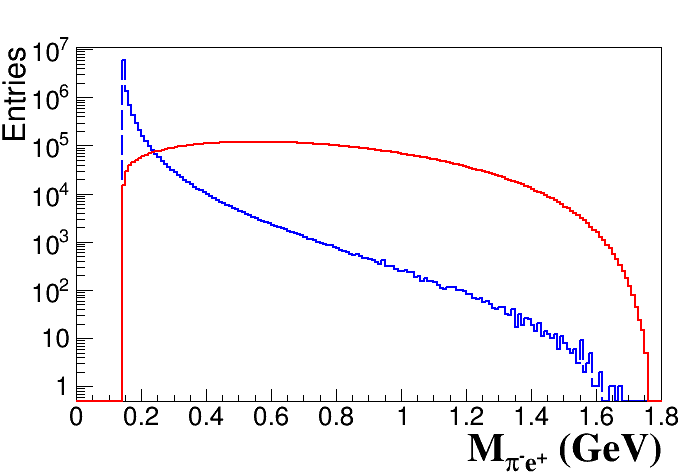}
\caption{Example plots from {\tt MC-TESTER}~\cite{Golonka:2002rz, Davidson:2008ma}: 
comparison for
$\tau^-\to\pi^- e^-e^+ \nu_{\tau}$ decay.
Dashed blue line represents model of Eq.~\eqref{me2} and
solid red line models Eq.~\eqref{me1}. 
Invariant masses of $e^+e^-$, $\pi^-e^+e^-$,
$\pi^-e^-$ and $\pi^-e^+$ systems are shown on 
top-right, top-left, bottom-left and bottom-right, respectively.
We can see that extended matrix element resulted in (as expected)
peaks in invariant mass of $\pi^- e^-e^+$ and $\pi^- e^+$, $\pi^- e^-$ systems.
Samples of 10M events were used for this comparison.}
\label{piee-ext} 
\end{figure}

\begin{figure}
\includegraphics[width=0.5\textwidth]{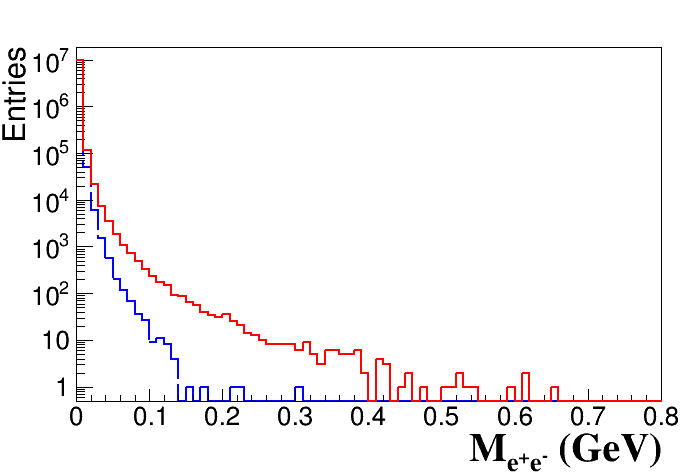}
\includegraphics[width=0.5\textwidth]{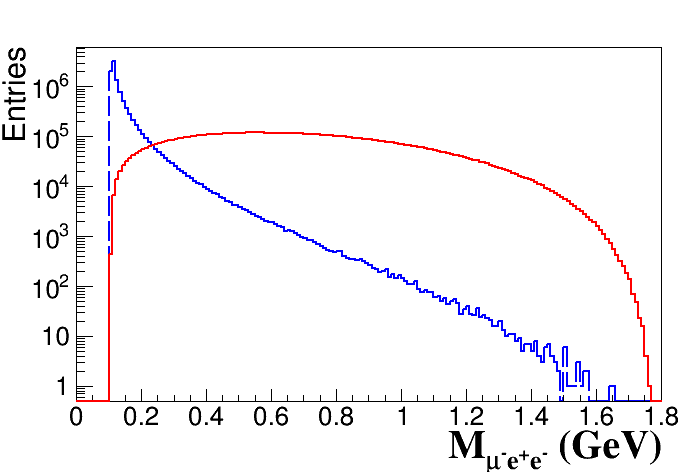}
\includegraphics[width=0.5\textwidth]{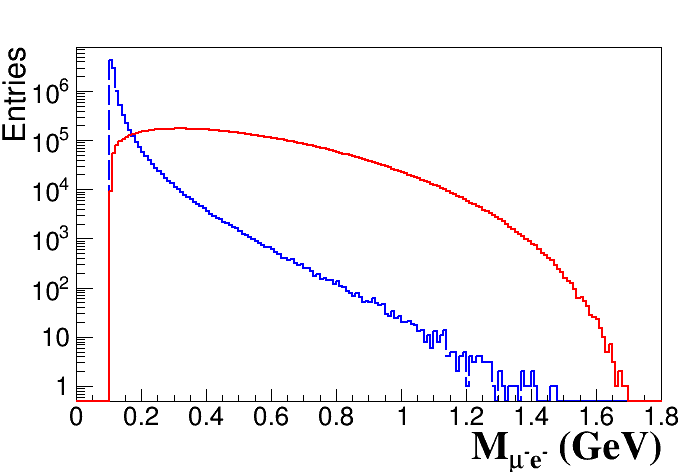}
\includegraphics[width=0.5\textwidth]{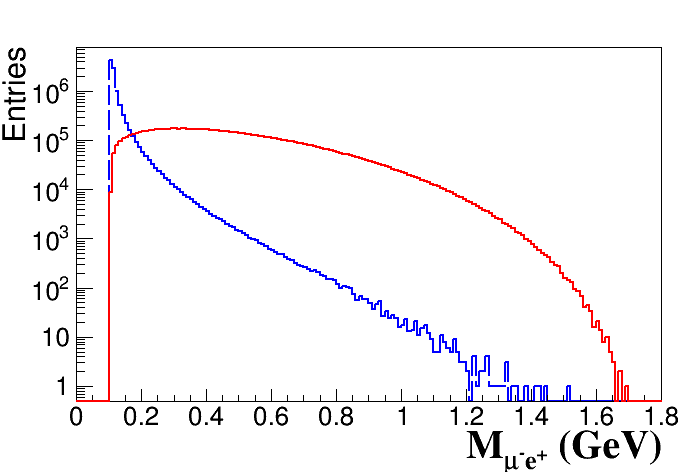}
\caption{Example plots from {\tt MC-TESTER}~\cite{Golonka:2002rz, Davidson:2008ma}: 
comparison for 
$\tau^-\to\bar{\nu}_\mu\mu^- e^-e^+\nu_\tau$ decay.
Dashed blue (light grey) line represents model of Eq.~\eqref{me2} and
solid red (dark grey) line models Eq.~\eqref{me1}.
Invariant masses of $e^+e^-$, $\mu^-e^+e^-$,
$\mu^-e^-$ and $\mu^-e^+$ systems are shown on 
top-right, top-left, bottom-left and bottom-right, respectively.
We can see that extended matrix element resulted in (as expected)
peaks in invariant mass of $\mu^- e^-e^+$ and $\mu^- e^+$, $\mu^- e^-$ systems.
Samples of 10M events were used for this comparison.}
\label{muee-ext}
\end{figure}

The demo model of Eq.~\eqref{me1} emulates the most important feature of the models containing $e^+e^-$ pair of the final state, which manifests as a soft singularity.
Eq.~\eqref{me2} emulates in addition a secondary feature, e.g.
collinear singularity of the emitted pair.
In terms of the invariant masses,
this should result in narrow peak in the mass of
three charged particles of the final state.
Fig.~\ref{piee-ext} and Fig.~\ref{muee-ext},
compare the distributions obtained
with models of Eq.~\eqref{me1} and Eq.~\eqref{me2}
for $\tau^-\to\pi^- e^-e^+ \nu_{\tau}$ 
and $\tau^-\to\bar{\nu}_\mu\mu^- e^-e^+\nu_\tau$ decays,
respectively, and show that this is indeed the case.

As expected, generating events with new matrix element of Eq.~\eqref{me2}
brought back the problem of low efficiency of generation.
We concentrate on parameters for
invariant mass of three charged particles\footnote{
Note that phase space generator for 5 particles has
only resonant type phase space parameterization
in the mass of four particles ($\nu_\tau$ excluded).
By default parameters there are set to MR=1.251 and GR=0.599.
Leaving those parameters unmodified is 
appropriate for flat phase space. 
There is no need to add new presampler channel for the mass of 4 particles.
}, as this is where the new peak appears.

In Tab. \ref{pi-ext} we collect the results for
$\tau^-\to\pi^- e^-e^+\nu_\tau$ decay.
We can see there that some over-weighted events remain even for well optimized
phase space parameters, and we get over 30\% efficiency of generation, 
which is very much acceptable.

\begin{table}[h]
\begin{center}
\resizebox{\textwidth}{!}{
\begin{tabular}{|l|l|l|l|l|l|}
\hline
\multicolumn{2}{|c}{Presampler param.}&\multicolumn{4}{|c|}{ Presampler performance }\\
\hline
MX & GX & eff. & overw.
& width[GeV] & rel.error \\
\hline
0.9 & 0.8 & 0.001 & 0\% (12) & $3.2315$/NaN & 0.0052/NaN\\
\hline
0.2 & 0.2 & 0.029 & 0\% (6) & $3.7326$ & 0.0001\\
\hline
0.17 & 0.1 & 0.068 & 0\% (5) & $3.7316$ & 0.0001\\
\hline
0.15 & 0.05 & 0.190 & 0\% (6) & $3.7315$ & 0.0001\\
\hline
0.13 & 0.05 & 0.305 & 0\% (9) & $3.7317$ & 0.0001\\
\hline
\end{tabular}}
\end{center}
\caption{Presampler parameters, resulting efficiency of generation,
over-weighted events, width and relative error of the width
for demo model of $\tau-\to\pi^- e^-e^+\nu_\tau$ decay.
Sample size of 10M events was used. In case of NaN obtained with 10M sample,
values obtained with 10k sample are given. }
\label{pi-ext}
\end{table}

In Tab. \ref{mu-ext} we collect the results for
$\tau-\to \bar{\nu}_\mu \mu^- e^-e^+\nu_\tau$ decay.
We can see there that  efficiency of generation is somewhat poor (~6\%).
This should not trouble us too much as channels with higher multiplicity of
final state particles tend to have lower efficiency of generation. 
For more complicated models, the efficiency should drop even more.

\begin{table}[h]
\begin{center}
\resizebox{\textwidth}{!}{
\begin{tabular}{|l|l|l|l|l|l|l|}
\hline
\multicolumn{3}{|c}{Presampler param.}&\multicolumn{4}{|c|}{ Presampler performance }\\
\hline
$P_A$ & MA & GA & eff. & overw.
& width[GeV] & rel.error \\
\hline
0.0 & 0.7759 & 0.1479 & 0.003 & 8\% (886001) & $0.3230\cdot10^{-1}$ & 0.0001\\
\hline
0.6 & 0.17 & 0.1 & 0.003 & 0\% (0) & $0.3230\cdot10^{-1}$ & 0.0001\\
\hline
0.7 & 0.15 & 0.05 & 0.015 & 0\% (0) & $0.3231\cdot10^{-1}$ & 0.0001\\
\hline
0.8 & 0.105 & 0.005 & 0.060 & 0\% (0) & $0.3231\cdot10^{-1}$ & 0.0001\\
\hline
\end{tabular}}
\end{center}
\caption{Presampler parameters, resulting efficiency of generation,
over-weighted events, width and relative error of the width
for demo model of $\tau-\to\bar{\nu}_\mu \mu^- e^-e^+\nu_\tau$ decay.
Sample size of 10M events was used. For results documented here we used MR=1.0 and GR=0.9.}
\label{mu-ext}
\end{table}

\section{Phase space presamplers tests}\label{doublep}

With the introduction of the new channels featuring $e^+e^-$ pair
(as well as channels with photon) in the final state,
the question of numerical precision arose.
Similar to the case of {\tt PHOTOS} Monte Carlo~\cite{Davidson:2010ew},
double precision accuracy is now necessary for generation of phase space in {\tt TAUOLA}.

There are 6 phase space generators used for each group of decay channels
with a specific amount of final state particles.
These are generators for 2, 3, 4, 5, 6 and N particles,
where the last one supports up to 9 particles.
Special generator introduced for 
$\tau \to \ell \nu_\ell \bar{\nu}_\tau (\gamma)$ decays
does not need to be modified by the user.
As the arguments of routines
and parameters in the library uses float precision,
we chose a solution that maintains backward compatibility.
Arguments for phase space generation routines remain float, 
while all the calculations within, are performed in double precision.
Such procedure guaranties that any models developed 
for the previous {\tt TAUOLA} release work right away, without any modifications. 

\begin{figure}
\includegraphics[width=0.5\textwidth]{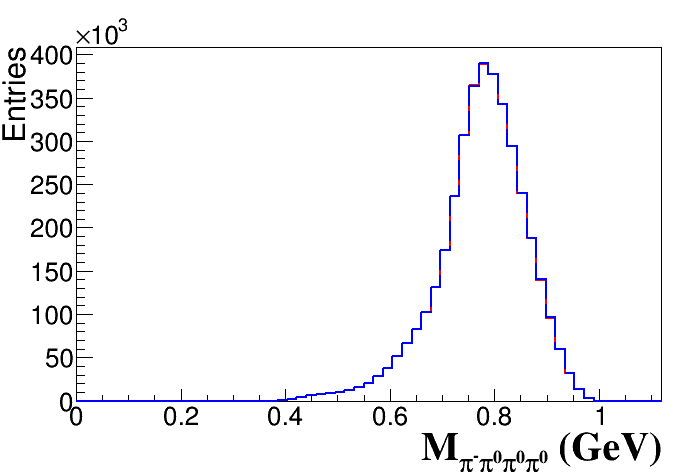}
\includegraphics[width=0.5\textwidth]{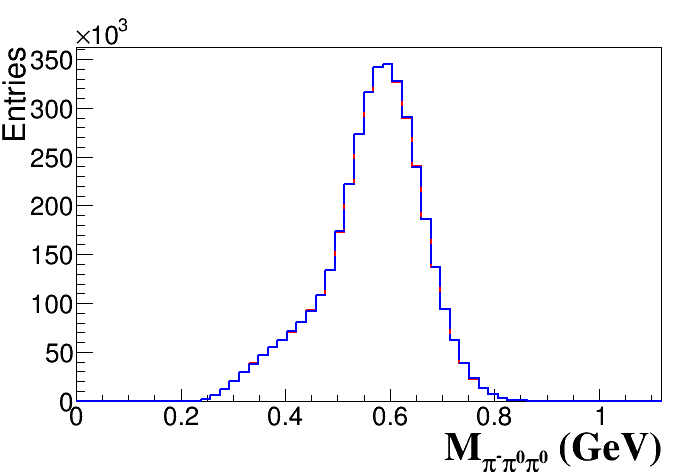}
\includegraphics[width=0.5\textwidth]{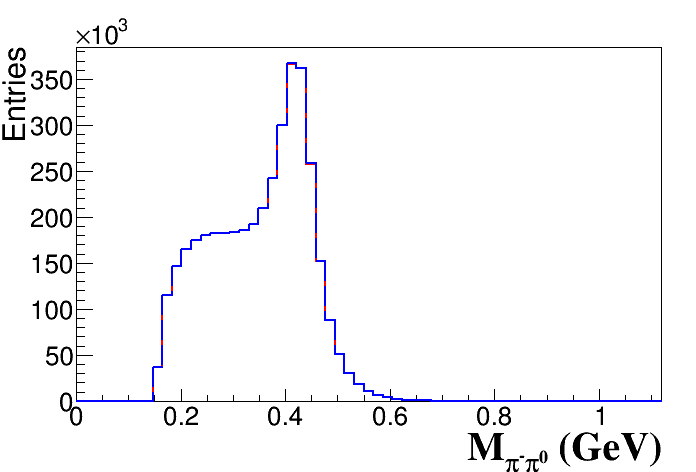}
\includegraphics[width=0.5\textwidth]{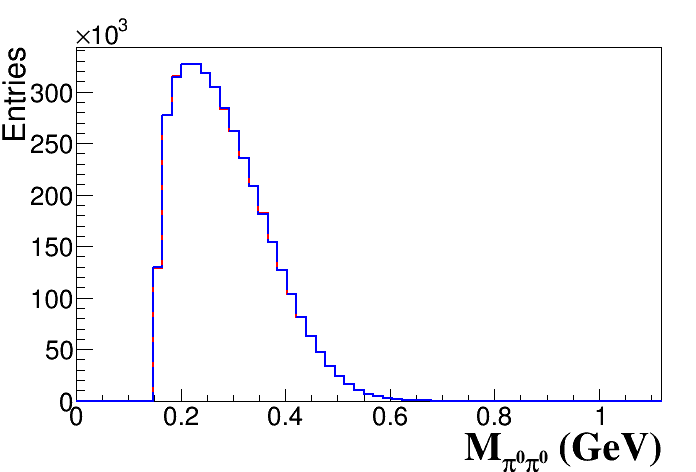}
\caption{Example plots from {\tt MC-TESTER}~\cite{Golonka:2002rz, Davidson:2008ma}: 
comparison of channels that could be affected by
changes introduced into phase space generator for decays into 5 particles.
Dashed blue lines represent the plots before changes 
and the solid red lines model the shapes after introduction of double precision.
Shown are selected plots for $\tau^-\to\pi^-\pi^0\pi^0\pi^0\nu_\tau$ 
decay channel with the original and updated code. 
Invariant masses of $\pi^-\pi^0\pi^0\pi^0$, $\pi^-\pi^0\pi^0$,
$\pi^-\pi^0$ and $\pi^0\pi^0$ systems are shown on 
top-right, top-left, bottom-left and bottom-right, respectively.
Full comparison contains dozens of similar plots and 
features SDP=0 (see Section 4 of~\cite{Golonka:2002rz}) for all of them.
}
\label{inny}
\end{figure}

For validation of the changes we have used
{\tt MC-TESTER} to confirm that no technical problem arises in any of 
decay channels present in {\tt TAUOLA} due to updates from this paper.
The Shape Difference Parameter (SDP),
for exact definition see Section 4 of~\cite{Golonka:2002rz}, 
for all distributions are equal to 0.
This was tested for each channel 
separately, for groups of channels, 
as well as for a run where all the
channels are generated simultaneously,
with sample sizes at least of 10M events for each run.
As an example, Fig.~\ref{inny} from tests of 
$\tau$ decay to 5 particles are presented,
using a generated sample of 100M events,
which gives almost 4M events for each decay channel of similar type.
The whole test consists of hundreds of histograms, 
while the upgrade to double precision itself is quite straightforward.
The compared distributions are nearly indistinguishable,
and until now such checks were not fundamental, the ones of many years ago were sufficient.
But with improved precision and peaked matrix elements as of the present paper,
they need to be repeated.

\section{Implementation example}\label{exampl}

Ref.~\cite{Guevara:2013wwa} (Eq. 31) provides formula 
for bremsstrahlung of pair emission which has been
introduced for physical modeling for pair emission into {\tt TAUOLA}.
Implementation can thus be cross-checked with analytical results. 

For the case of $\tau^-\to\pi^-e^+e^-\nu_\tau$ decays, Ref.~\cite{Guevara:2013wwa}
predicts branching ratio of $(1.461 \pm 0.0006) \times 10^{-5}$,
the Belle collaboration reports~\cite{Jin:2019goo} branching ratio between 
$(1.46\pm0.13\pm0.21)\times 10^{-5}$ and $(3.01\pm0.27\pm0.43)\times 10^{-5}$
for the structure-dependent mechanisms mediated by axial-vector and vector currents,
while our MC simulation produces a result of
$(1.398 \pm 0.001) \times 10^{-5}$.
In case of $\tau^-\to\pi^-\mu^+\mu^-\nu_\tau$, Ref.~\cite{Guevara:2013wwa}
gives branching ratio of $(1.6 \pm 0.007) \times 10^{-7}$,
while the Belle collaborations quotes an upper limit
of $1.14 \times 10^{-5}$ at 90\% confidence level.
Our MC simulation produces 
$(1.035 \pm 0.003)\times 10^{-7}$.
Those results seem to be in reasonable agreement and are collected in
Table~\ref{data1}. 

\begin{table}[h]
\begin{center}
\resizebox{\textwidth}{!}{
\begin{tabular}{|l|l|l|l|}
\hline
Decay mode & theoretical BR & Monte Carlo BR & experimental BR \\
\hline
$\tau^-\to\pi^-e^+e^-\nu_\tau$ & $(1.461 \pm 0.006) \times 10^{-5}$ & $(1.396 \pm 0.0004) \times 10^{-5}$ & <$3.01^{+0.27+0.43}_{-0.27-0.43}\times 10^{-5}$ \\
\hline
$\tau^-\to\pi^-\mu^+\mu^-\nu_\tau$ & $(1.6 \pm 0.007) \times 10^{-7}$ & $(1.038 \pm 0.0003)\times 10^{-7}$ & <$1.14\times 10^{-5}$ \\
\hline
\end{tabular}}
\end{center}
\caption{Comparison of branching ratios obtained from our MC implementation of model based on Ref.~\cite{Guevara:2013wwa} 
with theoretical prediction (from the above-mentioned reference) and experimental results from the Belle experiment~\cite{Jin:2019goo}.}
\label{data1}
\end{table}

\begin{figure}
\includegraphics[width=0.5\textwidth]{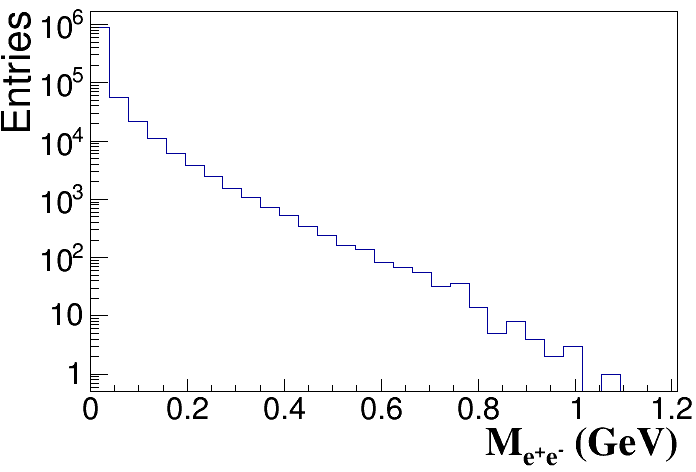}
\includegraphics[width=0.5\textwidth]{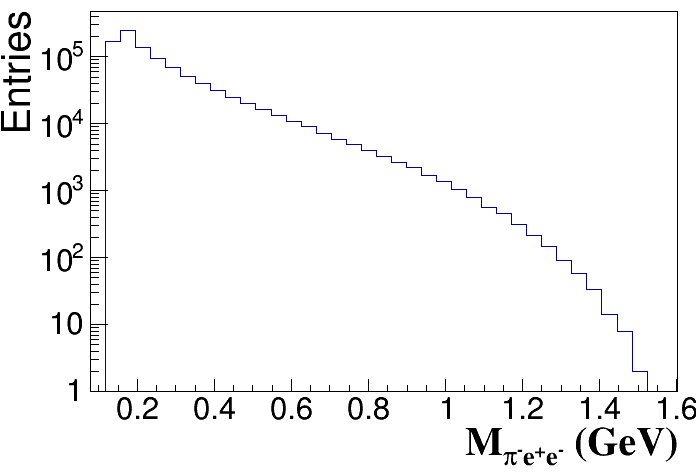}
\includegraphics[width=0.5\textwidth]{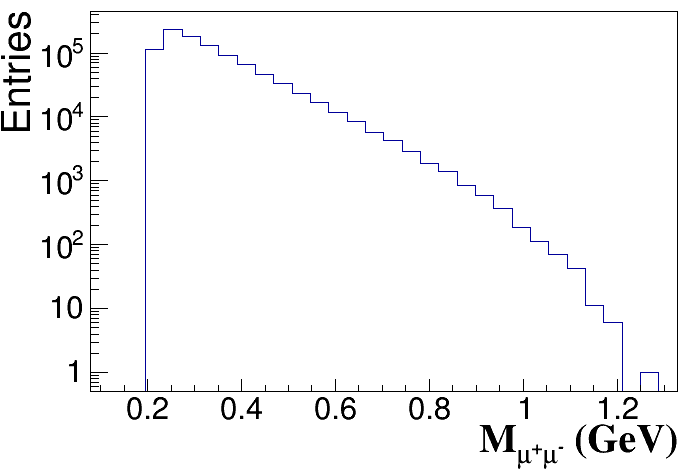}
\includegraphics[width=0.5\textwidth]{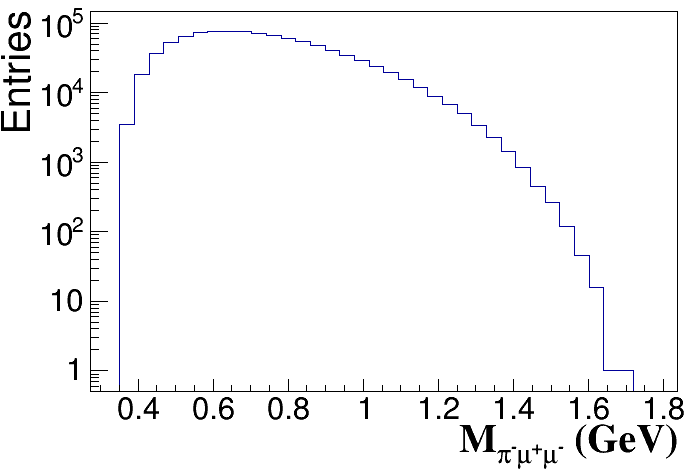}
\caption{
Invariant masses of $e^+e^-$ and $\pi^-e^+e^-$ systems 
in $\tau^-\to\pi^-e^+e^-\nu_\tau$ decays
are shown on top-left and top-right, and those of
$\mu^+\mu^-$ and $\pi^-\mu^+\mu^-$ systems 
in $\tau^-\to\pi^-\mu^+\mu^-\nu_\tau$ decays on 
bottom-left and bottom-right, respectively.
Plots obtained from MC samples of 4M events.}
\label{pipairs}
\end{figure}

There are multiple possible sources of the remaining differences:
phase space approximation in analytical calculations consistent with matrix element; 
exact phase space used in MC generation;
differences in parameters values; 
values of masses of intermediate particles, $|V_{ud}|$, etc.
Selected distributions of invariant mass  
obtained from our simulation are shown in Fig~\ref{pipairs}.

\begin{figure}
\includegraphics[width=0.5\textwidth]{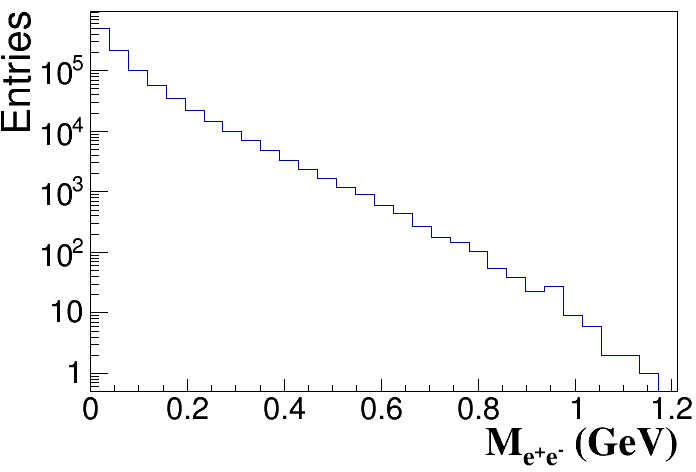}
\includegraphics[width=0.5\textwidth]{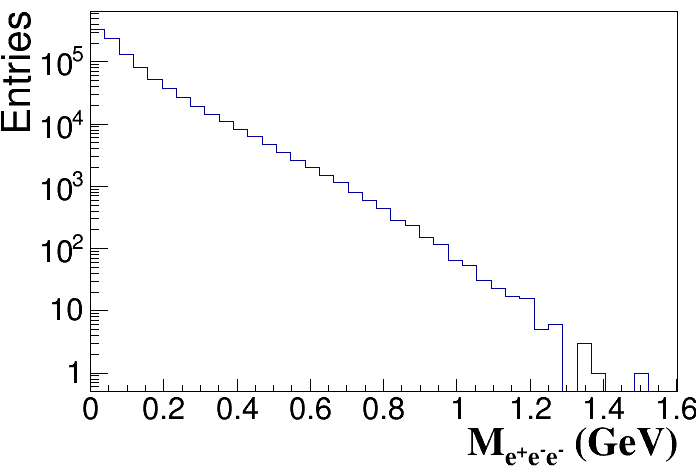}
\includegraphics[width=0.5\textwidth]{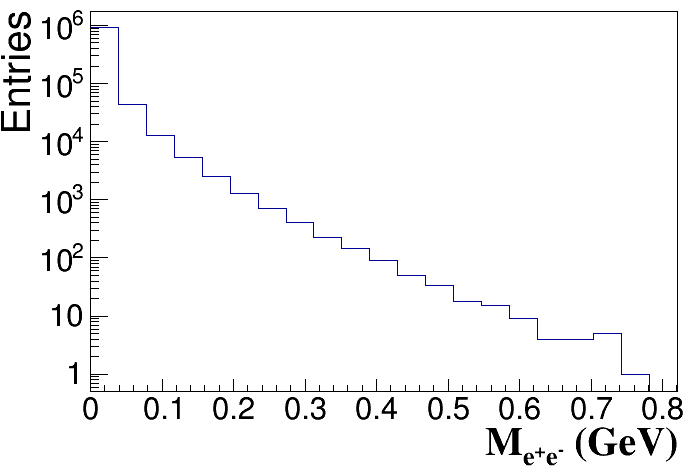}
\includegraphics[width=0.5\textwidth]{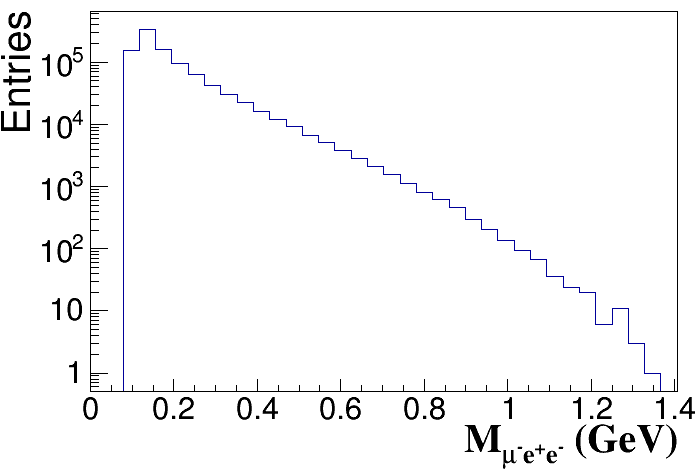}
\caption{
Invariant masses of $e^+e^-$ and $e^-e^+e^-$ systems 
in $\tau^-\to\bar{\nu}_e e^- e^+ e^-\nu_\tau$ decays
are shown on top-left and top-right, and those of
$e^+e^-$ and $\mu^-e^+e^-$ systems 
in $\tau^-\to\bar{\nu}_\mu \mu^-e^+e^-\nu_\tau$ decays on 
bottom-left and bottom-right, respectively.
In case of $\tau^-\to\bar{\nu}_e e^- e^+ e^-\nu_\tau$ decays,
the distributions of the invariant masses of the $e^+e^-$ systems
are symmetrized over the two possible final states.
Plots obtained from MC samples of 4M events.
}
\label{mupairs}
\end{figure}

In the case of $\tau^-\to\bar{\nu}_\mu\mu^- e^-e^+\nu_\tau$ and 
$\tau^-\to\bar{\nu}_e e^- e^-e^+\nu_\tau$ decays
we use matrix element squared described by the following factorized form:

{\small
\begin{equation}
|{\cal M}|^2=|{\cal M(\tau \to \ell \nu_\ell \nu_\tau)}|^2 \cdot 
2e^{4}\frac{4p_{e-}^{\alpha}p_{e+}^{\beta}-k^{2}g^{\alpha\beta}}{k^{4}}\left(\frac{p_{\ell}}{kp_{\ell}}-\frac{p_{\tau}}{kp_{\tau}}\right)_{\alpha}\left(\frac{p_{\ell}}{kp_{\ell}}-\frac{p_{\tau}}{kp_{\tau}}\right)_{\beta},
\label{fullme}
\end{equation}}
where {$\ell$} denotes a light lepton (electron or muon) and 
$\alpha$, $\beta$ are Lorentz indices.
With this we have run multiple technical tests, 
details of which are described in Appendix~\ref{sergiej}.
We conclude that with such a physical model\footnote{
We assume soft pairs, 
therefore we can start from the matrix element for leptonic
$\tau$ decays and multiply it by a factor for pair emission.}, 
the results are in reasonable agreement with those obtained by
the CLEO experiment~\cite{Alam:1995mt}. We collect the results
in Table~\ref{data2} along with the values from theoretical
calculations published in Ref.~\cite{Flores-Tlalpa:2015vga}.
Selected distributions of invariant mass 
obtained from our simulation are shown in Fig~\ref{mupairs}.

\begin{table}[h]
\begin{center}
\resizebox{\textwidth}{!}{
\begin{tabular}{|l|l|l|l|}
\hline
Decay mode & theoretical BR & Monte Carlo BR & experimental BR \\
\hline
$\tau^-\to\bar{\nu}_e e^- e^-e^+\nu_\tau$ & $(4.21 \pm 0.01) \times 10^{-5}$ & $(2.871 \pm 0.0004)\times 10^{-5}$  & $2.7 _{-1.1 -0.4 -0.3} ^{+1.5 +0.4 +0.1}\times  10^{-5}$\\
\hline
$\tau^-\to\bar{\nu}_\mu\mu^- e^-e^+\nu_\tau$ & $(1.984 \pm 0.004) \times 10^{-5}$ & $(1.538 \pm 0.0005 )\times 10^{-5}$ & <$3.2\times 10^{-5}$ (at 90\% C.L.) \\
\hline
\end{tabular}}
\end{center}
\caption{Comparison of branching ratios obtained from MC implementation 
our model~\eqref{fullme} 
with theoretical prediction from Ref.~\cite{Flores-Tlalpa:2015vga} and experimental results from the CLEO experiment~\cite{Alam:1995mt}. }
\label{data2} 
\end{table}

We note that some of the numerical values of the input parameters,
such as meson decay constants and quark-mixing matrix elements, etc.
differ between the ones used in Ref.~\cite{Guevara:2013wwa,Flores-Tlalpa:2015vga}
and ours, and as a consequence the numerical results slightly differ as well.

\section{Summary}\label{sum}

In this paper we document the motivation for a new {\tt TAUOLA} release,
and its modifications. The main purpose of this new version is to facilitate
models for generation of previously unconsidered features in phase space
arising from logarithmic peaks of infrared and acollinear nature in
bremsstrahlung-like emission, as opposed to standard Breit-Wigner ones. 
This is necessary to properly introduce models that study 
the background from $\tau$ decays in searches of New Physics, where $e^+e^-$
pair is produced through narrowly peaked virtual photon or dark scalar.
New phase space presamplers, to be used simultaneously with the already previously available,
for 4 and 5 particle final states were introduced.
Real variables were updated to double precision in all phase space generators.
Thanks to these changes, phase space generators are now suitable for particular class
of New Physics decay channels containing dark particles
that are modeled as narrow resonances decaying into lepton pair.
We leave corresponding discussion to a separate physics oriented publication~\cite{Banerjee:2021rtn}. 
On the other hand, we have checked already that additional technical changes will not be needed. 

Lepton  pairs close to $\tau$ lepton directions may originate from the $\tau$ production process.
If needed for background studies, they may  be generated,
e.g. with the help of {\tt PHOTOS} Monte Carlo~\cite{Davidson:2010ew}.
Also in case of $\tau$ lepton decays, additional lepton pairs may be generated with {\tt PHOTOS}.
Work on discussion of related systematic error has been pursued in Ref.~\cite{Banerjee:2021rtn}.

We provide also a more complete documentation for {\tt TAUOLA-bbb}~\cite{Chrzaszcz:2016fte}
in terms of the phase space presamplers.
This is for convenience to the user community to exploit all features of this tool.
Finally, we document new tests of {\tt TAUOLA} and its interface for user-defined models.
Examples of matrix elements based on Ref.~\cite{Guevara:2013wwa}
along with analytical and numerical calculations are confronted and presented.
More technical aspects and supplementary tests are delegated to Appendices. \\
For details on how to run the examples, Ref.~\cite{Chrzaszcz:2016fte}
and some explanations in the distribution tar ball can be used.

\centerline {\bf Acknowledgments}

This project was supported in part from funds of Polish National Science
Centre under decisions DEC-2017/27/B/ST2/01391,
and the U.S. Department of Energy and research Award No. DE-SC0022350.

\bibliographystyle{elsarticle-num}
\bibliography{references}

\appendix

\section{Phase space parameters and presamplers}
\label{sec:presamplers}
In this Appendix we describe the presampler parameters and how  the
phase space channels are prepared for each group of decays with distinct
number of particles in the final state.

Firstly, we would like to point, that
ordering of particles is important for  
proper functioning of the presampler. 
Therefore, the user should make sure that
resonances in the matrix element are
generated from four momenta of particles
that can be constructed with resonant type phase space.
Otherwise, the generation will suffer from extremely low efficiency or in severe cases generate incorrect distributions.
Except for channels with neutrino-less decays, 
the last particle should be always set $\nu_\tau$.
Neutrino-less decays can only be applied to channels with
2 or 3 particles in the final state.

Secondly, as mentioned in Section~\ref{phstruct},
it is important to note that although the phase space generators have multiple presampler channels,
not all of them are always used. 
Depending on the probabilities in their parameterization, 
specific $\tau$ decay channels can be activated.

In the following subsections, we will use the notation of $M^2_{abc}$, for example, to mean the invariant mass squared of system of 3 particles: a, b and c, respectively, where a, b and c are numbers given while ordering them accordingly. 

\subsection{Phase space for decays into 6 particles}\label{p6}
Presampler for decays into 6 particles has up to 4 channels.
Both, resonant and flat type phase space are 
present only for $M^2_{12345}$
and $M^2_{234}$. The use of this type of parameterization
is chosen randomly and independently of each other.
Other invariant masses are generated using only the flat type presampler.
Therefore, possible presampler channels are:
\begin{itemize}
 \item flat $M^2_{12345}$ $\to$ flat $M^2_{1234}$ $\to$ 
 flat $M^2_{234}$ $\to$ flat $M^2_{34}$,
 \item resonant $M^2_{12345}$ $\to$ flat $M^2_{1234}$ $\to$ 
 flat $M^2_{234}$ $\to$ flat $M^2_{34}$,
  \item flat $M^2_{12345}$ $\to$ flat $M^2_{1234}$ $\to$ 
 resonant $M^2_{234}$ $\to$ flat $M^2_{34}$,
  \item resonant $M^2_{12345}$ $\to$ flat $M^2_{1234}$ $\to$ 
 resonant $M^2_{234}$ $\to$ flat $M^2_{34}$.
\end{itemize}

The presampler parameters are:
\begin{itemize}
 \item $P_A$ - probability of resonant type phase space in $M^2_{12345}$,
 \item $P_B$ - probability of resonant type phase space in $M^2_{234}$,
 \item MA - mass-like parameter for $M^2_{12345}$,
 \item GA - width-like parameter for $M^2_{12345}$,
 \item MB - mass-like parameter for $M^2_{234}$,
 \item GB - width-like parameter for $M^2_{234}$.
\end{itemize}

\subsection{Phase space for decays into 5 particles}\label{p5}
Presampler for $\tau$ decays into 5 particles has up to 6 channels.
Flat and resonant type phase space parameterization are present
in the mass of three particles excluding $\nu_{\tau}$.
The system is symmetrized for same type of particles
in the first and second place giving 3 options here:
flat parameterization in the invariant mass of three particles,
resonant parameterization in the $M^2_{234}$ and
resonant parameterization in the $M^2_{134}$.
If first and second particles are different,
the latter generation channel is discarded.
Resonant type phase space is always used for the $M^2_{1234}$.
Also, there are internally defined parameters for resonant type
parameterization in the $M^2_{34}$ (see section~\ref{enhance}).
Therefore, possible presampler channels are:
\begin{itemize}
 \item resonant $M^2_{1234}$ $\to$ 
 flat $M^2_{234}$ $\to$ flat $M^2_{34}$,
 \item resonant $M^2_{1234}$ $\to$ 
 flat $M^2_{234}$ $\to$ resonant $M^2_{34}$,
  \item resonant $M^2_{1234}$ $\to$ 
 resonant $M^2_{234}$ $\to$ flat $M^2_{34}$,
  \item resonant  $M^2_{1234}$ $\to$ 
 resonant $M^2_{234}$ $\to$ resonant $M^2_{34}$,
   \item resonant $M^2_{1234}$ $\to$ 
 resonant $M^2_{134}$ $\to$ flat $M^2_{34}$,
  \item resonant  $M^2_{1234}$ $\to$ 
 resonant $M^2_{134}$ $\to$ resonant $M^2_{34}$.
\end{itemize}

The presampler parameters available through user interface are:
\begin{itemize}
 \item $P_A$ - probability of resonant type phase space in $M^2_{234}$
 and $M^2_{134}$,
 \item $P_B$ - redundant parameter with same meaning and $P_A$,
 \item MR - mass-like parameter for $M^2_{1234}$,
 \item GR - width-like parameter for $M^2_{1234}$,
 \item MA - mass-like parameter for $M^2_{234}$ and $M^2_{134}$,
 \item GA - width-like parameter for $M^2_{234}$ and $M^2_{134}$,
\end{itemize}
Probabilities $P_A$ and $P_B$ are summed up and the result is a probability of using
resonant type phase space parameterization in the mass of three particles.
Such structure was adopted due to the possible introduction of
additional generation channel in the mass of three particles
(mirroring other phase space generators),
but the need to fully exploit this feature never arose.
Then, $P_A$ and $P_B$ would have same 
function as in other phase space generators.
Internally defined parameters are: 
$M_{34}=0.001$, $\Gamma_{34}=0.001$ and $P_K$,
where $P_K$ is probability of resonant type presampler set to 0.8 if the $3^{rd}$ and the $4^{th}$ particles are electron and positron, respectively. 
Otherwise it is set to 0.0, 
which means only flat type phase space in present in $M_{34}^2$.

\subsection{Phase space for decays into 4 particles}\label{p4}
In case of channels with 4 particles in the final state, 
presampler has up to 3 channels.
All the three channels use resonant type phase space parameterization for $M^2_{123}$.
Channels A and B use parameterization for
$M^2_{13}$ or $M^2_{23}$, respectively,
but with different parameters,
while the third channel is uses flat type phase space
in the invariant mass squared of two particles.
Therefore, the possible presampler channels are:
\begin{itemize}
  \item  resonant $M^2_{123}$ $\to$ resonant $M^2_{13}$ (channel A),
  \item  resonant $M^2_{123}$ $\to$ resonant $M^2_{23}$ (channel B),
  \item  resonant $M^2_{123}$ $\to$ flat $M^2_{23}$.
\end{itemize}

The presampler has following 8 parameters:
\begin{itemize}
 \item $P_A$ - probability of channel A,
 \item $P_B$ - probability of channel B,
 \item MX - mass-like parameter for $M^2_{123}$,
 \item GX - width-like parameter for $M^2_{123}$,
 \item MA - mass-like parameter for channel A,
 \item GA - width-like parameter for channel A,
 \item MB - mass-like parameter for channel B,
 \item GB - width-like parameter for channel B.
\end{itemize}
The probability of channel with flat phase space in the mass of two particles equals $1-P_A-P_B$.
If $P_A=P_B=0.0$, only the third channel is used.

\subsection{Phase space for decays into 3 particles}\label{p3}
The presampler for decays into 3 particles has 3 channels,
all of which differ only in the phase space parameterization used for the $M^2_{12}$. 
Therefore, possible presampler channels are:
\begin{itemize}
  \item  resonant $M^2_{12}$ (channel A),
  \item  resonant $M^2_{12}$ (channel B),
  \item  flat $M^2_{12}$.
\end{itemize}

The presampler has following parameters:
\begin{itemize}
 \item $P_A$ - probability of channel A,
 \item $P_B$ - probability of channel B,
 \item MA - mass-like parameter for channel A,
 \item GA - width-like parameter for channel A,
 \item MB - mass-like parameter for channel B,
 \item GB - width-like parameter for channel B.
\end{itemize}
Probability of flat phase space (third channel) in the $M^2_{12}$ equals $1-P_A-P_B$.

\subsection{Phase space for decays into N and 2 particles}\label{pn}

Presamplers for decays into 2 and N particles do not have any parameters.
Decay into 2 particles does not need parameters for obvious reason. Presampler for N particles
can be used for up to 9 particles in final state but always
uses flat phase space for invariant mass squared 
of every system with descending number of particles e.g. for N=9: $M^2_{12345678}, M^2_{2345678}, ... , M^2_{678}, M^2_{78}$.
Use of matrix element is restricted.

\section{Technical tests for phase space and pair emission in 
$\tau^- \to \bar{\nu}_\mu \mu^- \nu_{\tau}$}\label{sergiej}

In this appendix we collect the formulae which 
numerical algorithm of {\tt TAUOLA} \cite{Jadach:1993hs} relies on.
They also provide platform to perform tests. 
We focus on a pair of channels 
$\tau^- \to \bar{\nu}_\mu \mu^-         \nu_{\tau}$ and 
$\tau^- \to \bar{\nu}_\mu \mu^- e^- e^+ \nu_{\tau}$, 
but obtained formulae are of use for 
$\tau^- \to \bar{\nu}_e e^-         \nu_{\tau}$ and 
$\tau^- \to \bar{\nu}_e e^- e^- e^+ \nu_{\tau}$ channels as well. 
The second channel in each pair differs from the first by the presence of a $e^-e^+$ pair and can be understood as a contribution to bremsstrahlung correction.
The dominant contribution is due to $e^-e^+$ pair of small virtuality (originating from the decay of nearly real photon).
In calculations we use in general notation of~\cite{Was:1994kg}. 
We shorten: 
$\bar{\nu}$ mean $\bar{\nu}_\ell$, where $\ell$, either electron or muon, 
and $\nu$ means $\nu_\tau$.

\vskip 2mm
\centerline{\bf 3 body decay}
\vskip 2mm

An integral of matrix element squared 
$\left|M\right|^{2}\equiv\left|M(p_{\tau},p_{\nu},p_{\overline{\nu}},p_{\mu})\right|^{2}$ over 3-body phase space $dLips_{3}(p_{\tau},p_{\nu},p_{\overline{\nu}},p_{\mu})$ reads:

\begin{align}
 & \resizebox{0.9\linewidth}{!}{%
 $\displaystyle
 \int\left|M\right|^{2}dLips_{3}(p_{\tau},p_{\nu},p_{\overline{\nu}},p_{\mu})=\int\left|M\right|^{2}\frac{d^{3}p_{\nu}}{(2\pi)^{3}2p_{\nu}^{0}}\frac{d^{3}p_{\overline{\nu}}}{(2\pi)^{3}2p_{\overline{\nu}}^{0}}\frac{d^{3}p_{\mu}}{(2\pi)^{3}2p_{\mu}^{0}}(2\pi)^{4}\delta^{4}(p_{\tau}-p_{\nu}-p_{\overline{\nu}}-p_{\mu})=$}\nonumber \\
 & \resizebox{0.8\linewidth}{!}{%
 $\displaystyle
 =\frac{1}{2^{11}\pi{}^{5}}\intop_{m_{\mu}^{2}}^{\left(m_{\tau}-m_{\mu}\right)^{2}}dM_{\overline{\nu}\mu}^{2}\intop_{-1}^{1}dcos\theta_{\nu}\intop_{0}^{2\pi}d\varphi_{\nu}\left(1-\frac{M_{\overline{\nu}\mu}^{2}}{m_{\tau}^{2}}\right)\intop_{-1}^{1}dcos\theta_{\overline{\nu}}\intop_{0}^{2\pi}d\varphi_{\overline{\nu}}\left(1-\frac{m_{\mu}^{2}}{M_{\overline{\nu}\mu}^{2}}\right)\left|M\right|^{2}$},\label{3body}
\end{align}
\noindent
where $p_{\tau}$, $p_{\nu}$, $p_{\overline{\nu}}$, $p_{\mu}$ are four-momenta of $\tau^-$, $\nu$, $\overline{\nu}_\mu$, $\mu^-$ correspondingly; $dcos\theta_{\nu}d\varphi_{\nu}$ is the solid angle element of $p_{\nu}$ in the rest frame of $\tau^-$, $dcos\theta_{\overline{\nu}}d\varphi_{\overline{\nu}}$ is the solid angle element of $p_{\overline{\nu}}$ in the rest frame of $(p_{\overline{\nu}}+p_{\mu})$; $M_{\overline{\nu}\mu}^2\nolinebreak=\nolinebreak(p_{\overline{\nu}}\nolinebreak+\nolinebreak p_{\mu})^2$; $m_{\mu}$ is mass of $\mu^-$ and $m_{\tau}$ is mass of $\tau^-$.

\vskip 2mm
\centerline{\bf 5 body decay}
\vskip 2mm

We proceed with writing a cross section for the 5-body decay
$\tau^{-}\to\bar{\nu}_{\mu}\mu^{-}e^{-}e^{+}\nu_{\tau}$
assuming the matrix element 
$\left|M\right|^{2}\equiv\left|M\left(p_{\tau},p_{e-},p_{e+},p_{\nu},p_{\overline{\nu}},p_{\mu}\right)\right|^{2}$
can be factorized. We focus on soft pair emissions:

\begin{equation}
\left|M\right|^{2}=\left|M\left(p_{\tau},p_{\nu},p_{\overline{\nu}},p_{\mu}\right)\right|^{2}\times\left|M_{F}\left(p_{e-},p_{e+}\right)\right|^{2}.
\label{fac}
\end{equation}
\noindent
Therefore:
\begin{align}
 & \int\left|M\right|^{2}dLips_{5}\left(p_{\tau},p_{e-},p_{e+},p_{\nu},p_{\overline{\nu}},p_{\mu}\right)= \nonumber\\
 & =\int\left|M_{F}\right|^{2}\frac{d^{3}p_{e-}}{(2\pi)^{3}2p_{e-}^{0}}\frac{d^{3}p_{e+}}{(2\pi)^{3}2p_{e+}^{0}}~d^{4}R~\delta^{4}(R-p_{\tau}+p_{e-}+p_{e+})\times \nonumber \\
 & \resizebox{0.9\linewidth}{!}{%
 $\displaystyle \times\int\left|M\left(p_{\tau},p_{\nu},p_{\overline{\nu}},p_{\mu}\right)\right|^{2}\frac{d^{3}p_{\nu}}{(2\pi)^{3}2p_{\nu}^{0}}\frac{d^{3}p_{\overline{\nu}}}{(2\pi)^{3}2p_{\overline{\nu}}^{0}}\frac{d^{3}p_{\mu}}{(2\pi)^{3}2p_{\mu}^{0}}(2\pi)^{4}\delta^{4}(R-p_{\nu}-p_{\overline{\nu}}-p_{\mu})$},
\end{align}
\noindent
where $p_{\tau}$, $p_{e-}$, $p_{e+}$, $p_{\nu}$, $p_{\overline{\nu}}$,
$p_{\mu}$ are four-momenta of $\tau^{-}$, $e^{-}$, $e^{+}$, $\nu$,
$\overline{\nu}_{\mu}$, $\mu^{-}$ correspondingly. At first and for a test, we put $\left|M_{F}\right|^{2}\equiv1$. Since the factorized part of matrix element squared $\left|M_{F}\right|^{2}$ does not depend
on $p_{e-}$, $p_{e+}$ anymore, for a soft pair emission we can drop
$e^{+}$ and $e^{-}$ from the conditions of momentum-energy conservation.
Thus the technical integral element 
$d^{4}R~\delta^{4}(R-p_{\tau}+p_{e-}+p_{e+})$ 
reduces to $R=p_{\tau}$ and

\begin{align}
 & \int\left|M\right|^{2}dLips_{5}\left(p_{\tau},p_{e-},p_{e+},p_{\nu},p_{\overline{\nu}},p_{\mu}\right)\approx \nonumber \\
 & =\int\frac{d^{3}p_{e-}}{(2\pi)^{3}2p_{e-}^{0}}\frac{d^{3}p_{e+}}{(2\pi)^{3}2p_{e+}^{0}}\int\left|M\right|^{2}dLips_{3}\left(p_{\tau},p_{\nu},p_{\overline{\nu}},p_{\mu}\right)= \nonumber \\
 & \resizebox{0.9\linewidth}{!}{%
 $\displaystyle =\frac{1}{2^{8}\pi{}^{6}}\int\left[dcos\theta_{e-}d\varphi_{e-}\frac{|\overline{p}_{e-}|^{2}d|\overline{p}_{e-}|}{\sqrt{|\overline{p}_{e-}|^{2}+m_{e}^{2}}}\right]_{\overline{p}_{\tau}=0}\left[dcos\theta_{e+}d\varphi_{e+}\frac{|\overline{p}_{e+}|^{2}d|\overline{p}_{e+}|}{\sqrt{|\overline{p}_{e+}|^{2}+m_{e}^{2}}}\right]_{\overline{p}_{\tau}=0}\times$}\nonumber \\
 & \resizebox{0.6\linewidth}{!}{%
 $\displaystyle \times\int\left|M\left(p_{\tau},p_{\nu},p_{\overline{\nu}},p_{\mu}\right)\right|^{2}dLips_{3}(p_{\tau},p_{\nu},p_{\overline{\nu}},p_{\mu})$},
\end{align}
\noindent
where $\overline{p}_{e-}$, $\overline{p}_{e+}$ are three-momenta
of $e^{-}$, $e^{+}$ correspondingly; subscript $\overline{p}_{\tau}=0$
or $\overline{p}_{\overline{\nu}}+\overline{p}_{\mu}=0$ means that
the variables into square brackets are in $\tau^{-}$ rest frame; $dcos\theta_{e-}d\varphi_{e-}$ is the solid angle
element of $p_{e^{-}}$, $dcos\theta_{e+}d\varphi_{e+}$ is the solid
angle element of $p_{e^{+}}$.

Our formula is valid for soft $e^+e^-$ only,
that is why we can work only with a part of the phase space.
We introduce a cutoff parameter $\Delta_1$: $p_{e+}^{0}<\Delta_1$,
$p_{e^{-}}^{0}<\Delta_1$. Such a conditions match the 
limitation introduced for the {\tt TAUOLA} generation. We obtain:

\begin{align}
 & \resizebox{0.9\linewidth}{!}{%
 $\displaystyle \int\left|M\right|^{2}dLips_{5}\left(p_{\tau},p_{e-},p_{e+},p_{\nu},p_{\overline{\nu}},p_{\mu}\right)\approx\frac{\Delta^{4}_1}{2^{6}\pi{}^{4}}\int\left|M\right|^{2}dLips_{3}\left(p_{\tau},p_{\nu},p_{\overline{\nu}},p_{\mu}\right)$}
\end{align}
and soft pair emission factor of the test reads:
\begin{align}
 B_{f}(\Delta_1)\approx\frac{\Delta^{4}_1}{2^{6}\pi{}^{4}}.\label{eq:bf1}
\end{align}
\noindent
Alternatively, the second test with $\left|M_{F}\right|^{2}\equiv1$
is to write cross section for the 5-body decay in terms of invariant
mass variables:

\begin{align}
 & \resizebox{0.9\linewidth}{!}{%
 $\displaystyle
 \int\left|M\right|^{2}dLips_{5}(p_{\tau})=\frac{1}{2^{11}\pi{}^{5}}\int dM_{\overline{\nu}\mu ee}^{2}\int d\Omega_{\nu}\left(1-\frac{M_{\overline{\nu}\mu ee}^{2}}{m_{\tau}^{2}}\right)\int d\Omega_{\overline{\nu}}\left(1-\frac{M_{\mu ee}^{2}}{M_{\overline{\nu}\mu ee}^{2}}\right)\left|M\left(p_{\tau},p_{\nu},p_{\overline{\nu}},p_{\mu}\right)\right|^{2}\times\label{eq:3body-02}$} \\
 & \resizebox{0.9\linewidth}{!}{%
 $\displaystyle
 \times\frac{1}{2^{12}\pi{}^{6}}\int d\Omega_{\mu}\int d\Omega_{e}\intop dM_{ee}^{2}\left|M_{F}\right|^{2}\sqrt{1-\frac{4m_{e}^{2}}{M_{ee}^{2}}}\intop dM_{\mu ee}^{2}\frac{\sqrt{\left(M_{\mu ee}^{2}-M_{ee}^{2}-m_{\mu}^{2}\right)^{2}-4M_{ee}^{2}m_{\mu}^{2}}}{M_{\mu ee}^{2}}$},\label{eq:factor-01}
\end{align}
\noindent
where $M_{\overline{\nu}\mu ee}^{2}=\left(p_{\overline{\nu}}+p_{\mu}+p_{e-}+p_{e+}\right)^{2}$,
$M_{\mu ee}^{2}=\left(p_{\mu}+p_{e-}+p_{e+}\right)^{2}$ , $M_{ee}^{2}=\left(p_{e-}+p_{e+}\right)^{2}$;
$d\Omega_{\nu}$ is the solid angle element of $p_{\nu}$ in the rest
frame of $\tau^{-}$, $d\Omega_{\overline{\nu}}$ is the solid angle
element of $p_{\overline{\nu}}$ in the rest frame of $\left(p_{e-}+p_{e+}+p_{\overline{\nu}}+p_{\mu}\right)$,
$d\Omega_{\mu}$ is the solid angle element of $p_{\mu}$ in the rest
frame of $\left(p_{e-}+p_{e+}+p_{\mu}\right)$, $d\Omega_{e}$ is
the solid angle element of $p_{e-}$ in the rest frame of $\left(p_{e-}+p_{e+}\right)$.
Considering pair emission is soft, we can approximate $M_{\overline{\nu}\mu ee}^{2}\approx M_{\overline{\nu}\mu}^{2}$,
$M_{\mu ee}^{2}=m_{\mu}^{2}$, thus first part of cross section \eqref{eq:3body-02}
coincide with cross section \eqref{3body} for 3-body decay $\tau^{-}\to\bar{\nu}_{\mu}\mu^{-}\nu_{\tau}$.
Soft pair emission factor reads:

\begin{align}
 & \resizebox{0.9\linewidth}{!}{%
 $\displaystyle
 B_{f}(\Delta_2)=\frac{1}{2^{8}\pi{}^{4}}\intop_{4m_{e}^{2}}^{\Delta^{2}_2}dM_{ee}^{2}\sqrt{1-\frac{4m_{e}^{2}}{M_{ee}^{2}}}\intop_{\left(m_{\mu}+M_{ee}\right)^{2}}^{\left(m_{\mu}+\Delta_2\right)^{2}}dM_{\mu ee}^{2}\frac{\sqrt{\left(M_{\mu ee}^{2}-M_{ee}^{2}-m_{\mu}^{2}\right)^{2}-4M_{ee}^{2}m_{\mu}^{2}}}{M_{\mu ee}^{2}}$}.\label{eq:factor-02}
\end{align}
\noindent
Here cutoff $\Delta_2$ limits invariant
mass of the $e^+e^-$ pair, therefore cutoff
could be invoked in {\tt TAUOLA} easily. 
Double integral of soft pair emission factor Eq.~\eqref{eq:factor-02}
doesn't have a simple analytical solution. 
On the other hand, numerical
solution works perfectly for testing purposes.

The $B_{f}$ of Eq.~\eqref{eq:factor-02} as function of $\Delta_2$ 
can be easily translated into $\Delta_2$ dependent partial widths 
simply multiplying partial width of 
$\tau^- \to \bar{\nu}_\mu \mu^-  \nu_{\tau}$ decay by $B_{f}(\Delta_2)$.
Results obtained that way and those from Monte Carlo simulation
are collected in Table~\ref{w1}. They provide also a test of 
approach used in Eq.~\eqref{fac} - tests with simplified matrix element. 

\begin{table}[h]
\begin{center}
\begin{tabular}{|l|l|l|}
\hline
\cline{1-3}\multicolumn{1}{|c|}{$\Delta_2$ [GeV]}
&\multicolumn{2}{|c|}{ Partial width [GeV]}\\
 \hline
  & $B_{f}(\Delta_2)\times\Gamma(\tau\to\bar{\nu}\mu\nu)$ & Monte Carlo \\
 \hline
 0.00125 & $0.42866\cdot 10^{-30}$ & $0.42729\cdot 10^{-30}$ \\
 \hline
 0.0025 & $0.16289\cdot 10^{-27}$ & $0.15965\cdot 10^{-27}$ \\
 \hline
 0.005 & $0.48627\cdot 10^{-26}$ & $0.46480\cdot 10^{-26}$ \\
 \hline
 0.01 & $0.92486\cdot 10^{-23}$ & $0.84837\cdot 10^{-23}$ \\
 \hline
 0.02 & $0.15208\cdot 10^{-22}$ & $0.12664\cdot 10^{-22}$ \\
 \hline
\end{tabular}
\end{center}
\caption{Partial width obtained for different cutoff $\Delta_2$
from Monte Carlo run and numerically from $B_{f}$ of Eq.\eqref{eq:factor-02}. 
Note, that with increasing cutoff $\Delta_2$, pairs are allowed to be
somewhat harder, therefore assumption $\Gamma(\tau\to\bar{\nu}\mu e e \nu)\approx B_f*\Gamma(\tau\to\bar{\nu}\mu\nu)$ . 
Uncertainty of MC results is at the level of 1\%.
}
\label{w1}
\end{table}

Similar tests with $\left|M_{F}\right|^{2}$ 
closer to a physical model can be also performed. 
Results for all such comparisons would be redundant, 
therefore we present only next step, 
where we choose factorized part of matrix element
squared to be: 
$\left|M_{F}\right|^{2}=\frac{\left(4\pi\alpha\right)^{2}}{\left(p_{e+}+p_{e-}\right)^{2}}$.
Such a choice should represent numerical effects of a singular behavior
during simulation of soft pair emission. Soft pair emission factor
in this case reads:

\begin{align}
 & \resizebox{0.9\linewidth}{!}{%
 $\displaystyle B_{f}=\int\left|M_{F}\right|^{2}\frac{d^{3}p_{e-}}{(2\pi)^{3}2p_{e-}^{0}}\frac{d^{3}p_{e+}}{(2\pi)^{3}2p_{e+}^{0}}d^{4}R\delta^{4}(R-p_{\tau}+p_{e-}+p_{e+})=$}\nonumber \\
 &
 =\frac{1}{(2\pi)^{6}}\int\frac{\left(4\pi\alpha\right)^{2}}{q^{2}}\frac{d^{3}p_{e-}}{2p_{e-}^{0}}\frac{d^{3}p_{e+}}{2p_{e+}^{0}}d^{4}qdM_{ee}^{2}\delta(q^{2}-M_{ee}^{2})\Theta(q^{0})\times\nonumber \\
 & \resizebox{0.9\linewidth}{!}{%
 $\displaystyle \times\delta^{4}(q-p_{e+}-p_{e-})d^{4}RdM_{\mu\nu\overline{\nu}}^{2}\delta(R^{2}-M_{\mu\nu\overline{\nu}}^{2})\Theta(R^{0})\delta^{4}(R-p_{\tau}+q)$},
\end{align}
\noindent
where we've introduced 
$q=p_{e-}+p_{e+}$; $R=p_{\nu}+p_{\overline{\nu}}+p_{\mu}$,
which represents 4-momentum of rest of the particles system after pair emission takes place,
and invariant mass squares $M_{ee}^{2}=\left(p_{e+}+p_{e-}\right)^{2}$ 
and 
$M_{\mu\nu\overline{\nu}}^{2}=\left(p_{\nu}+p_{\overline{\nu}}+p_{\mu}\right)^{2}$.
With help of formulae
\begin{align}
 & \int\frac{d^{3}p_{e-}}{2p_{e-}^{0}}\frac{d^{3}p_{e+}}{2p_{e+}^{0}}d^{4}q\delta^{4}(q-p_{e+}-p_{e-})=\int d^{4}qdcos\theta_{1}d\varphi_{1}\frac{1}{8}\sqrt{1-\frac{4m_{e}^{2}}{q^{2}}},\nonumber \\
 & \int d^{4}qdM_{ee}^{2}\delta(q^{2}-M_{ee}^{2})\Theta(q^{0})=\int\frac{d^{3}q}{2q^{0}},\nonumber \\
 & \int d^{4}RdM_{\mu\nu\overline{\nu}}^{2}\delta(R^{2}-M_{\mu\nu\overline{\nu}}^{2})\Theta(R^{0})=\int\frac{d^{3}R}{2R^{0}}
\end{align}
\noindent
$B_f$ reads:

\begin{align}
 B_{f}=\frac{\alpha^{2}}{2^{3}\pi^{3}}\int\frac{dM_{ee}^{2}}{M_{ee}^{4}}\sqrt{1-\frac{4m_{e}^{2}}{M_{ee}^{2}}}dM_{\mu\nu\overline{\nu}}^{2}dcos\theta_{2}d\varphi_{2}\frac{\lambda^{1/2}(m_{\tau}^{2},M_{ee}^{2},M_{\mu\nu\overline{\nu}}^{2})}{8m_{\tau}^{2}},
\end{align}
\noindent
where $\lambda^{1/2}(m_{\tau}^{2},M_{ee}^{2},M_{\mu\nu\overline{\nu}}^{2})=\sqrt{\left(m_{\tau}^{2}+M_{ee}^{2}-M_{\mu\nu\overline{\nu}}^{2}\right)^{2}-4m_{\tau}^{2}M_{\mu\nu\overline{\nu}}^{2}}$.

Integration over angular variables performs trivially. An easy way to proceed with integration is to integrate over energy 
$E_{ee}=\frac{m_{\tau}^{2}+M_{ee}^{2}-M_{\mu\nu\overline{\nu}}^{2}}{2m_{\tau}}$ of
the pair in the rest frame of $\tau^{-}$. For the condition $dM_{ee}^{2}=0$, 
differential of the energy of
the pair reads $dE_{ee}=-\frac{dM_{\mu\nu\overline{\nu}}^{2}}{2m_{\tau}}$ 
leading to $\lambda^{1/2}(m_{\tau}^{2},M_{ee}^{2},M_{\mu\nu\overline{\nu}}^{2})=2m_{\tau}\sqrt{E_{ee}^{2}-M_{ee}^{2}}$.

Soft pair emission factor reads:

\begin{align}
 & \resizebox{0.8\linewidth}{!}{%
 $\displaystyle B_{f}(\Delta_3)=\frac{\alpha^{2}}{2^{2}\pi^{2}}\intop_{4m_{e}^{2}}^{\Delta^{2}_3}\frac{dM_{ee}^{2}}{M_{ee}^{2}}\sqrt{1-\frac{4m_{e}^{2}}{M_{ee}^{2}}}\intop_{M_{ee}}^{\Delta_3}dE_{ee}\sqrt{E_{ee}^{2}-M_{ee}^{2}}=$} \nonumber \\
 & \resizebox{0.9\linewidth}{!}{%
 $\displaystyle =\frac{\alpha^{2}}{2^{3}\pi^{2}}\intop_{4m_{e}^{2}}^{\Delta^{2}_3}\frac{dM_{ee}^{2}}{M_{ee}^{2}}\sqrt{1-\frac{4m_{e}^{2}}{M_{ee}^{2}}}\left(\Delta_3\sqrt{\Delta^{2}_3-M_{ee}^{2}}+M_{ee}^{2}\ln\frac{M_{ee}}{\Delta_3+\sqrt{\Delta^{2}_3-M_{ee}^{2}}}\right)$},\label{eq:w2}
\end{align}
\noindent
where $\Delta_3$, maximal energy of the pair in the $\tau^{-}$ rest frame determines the $\Delta^2_3$ the maximal $M^2_{ee}$. Analytical
expression for $B_{f}(\Delta_3)$ for this choice of $\left|M_{F}\right|^{2}$
is long and is not instructive. 
Partial widths obtained using this new $B_{f}(\Delta_3)$ 
and those from Monte Carlo simulation
are collected in Table~\ref{w2}. 

\begin{table}[h]
\begin{center}
\begin{tabular}{|l|l|l|}
\hline
\cline{1-3}\multicolumn{1}{|c|}{$\Delta_3$ [GeV]}
&\multicolumn{2}{|c|}{ Partial width [GeV]}\\
 \hline
  & $B_{f}(\Delta_3)\times\Gamma(\tau\to\bar{\nu}\mu\nu)$ & Monte Carlo \\
 \hline
 0.0025 & $0.59970\cdot 10^{-24}$ & $0.59826\cdot 10^{-24}$ \\
 \hline
 0.005 & $0.82769\cdot 10^{-23}$ & $0.81803\cdot 10^{-23}$ \\
 \hline
 0.01 & $0.64485\cdot 10^{-22}$ & $0.64446\cdot 10^{-22}$ \\
 \hline
 0.02 & $0.39679\cdot 10^{-21}$ & $0.38269\cdot 10^{-21}$ \\
 \hline
\end{tabular}
\end{center}
\caption{Partial width obtained for different cutoff $\Delta_3$
from Monte Carlo run and numerically from $B_{f}$ of Eq.\eqref{eq:w2}. 
Note, that with increasing cutoff $\Delta_3$, 
pairs are allowed to be somewhat harder
therefore assumption $\Gamma(\tau\to\bar{\nu}\mu e e \nu)\approx B_f*\Gamma(\tau\to\bar{\nu}\mu\nu)$ 
works worse.
Uncertainty of MC results is at the level of 1\%.
Those results, together with the ones collected in Table~\ref{w1}, 
provide us with confirmation that approach of Eq.~\eqref{fac}
is justified as well as provide technical test 
for the phase space generation in {\tt TAUOLA}.}
\label{w2}
\end{table}

\end{document}